\documentclass[twocolumn]{aa}

\usepackage{color}
\usepackage{epsfig}
\usepackage{soul}
\usepackage{float}
\usepackage{enumitem}
\usepackage{appendix}
\usepackage{wrapfig}
\usepackage{graphicx}
\usepackage{multicol}
\usepackage{hyperref}
\usepackage{chemfig}
\usepackage{longtable}
\usepackage{supertabular,booktabs}
\usepackage{pdflscape}
\usepackage{longtable}
\usepackage{caption}
\definecolor{NGC1333}{HTML}{006D77}
\definecolor{B1}{HTML}{5C3D73}
\definecolor{IC348}{HTML}{922D50}
\definecolor{B5}{HTML}{DAA520}

\usepackage[version=3]{mhchem}
\hypersetup{
    colorlinks=true,
    linkcolor=blue,
    filecolor=magenta,
    citecolor=blue,
    urlcolor=blue,
    pdftitle={Overleaf Example},
    pdfpagemode=FullScreen,
    }
\raggedcolumns
\usepackage{natbib}

\def\aap{A\&A}
\def\apjs{ApJS}

\usepackage{blindtext}
\usepackage{lipsum}
\usepackage{afterpage}

\newsavebox{\splitcaptionbox}
\newlength{\splitcaptionheight}
\makeatletter
\long\def\@makesplitcaption#1#2{
  \if@stage@final
    \vskip0.7\abovecaptionskip
    \addtolength{\splitcaptionheight}{-0.7\abovecaptionskip}
  \else
    \vskip\abovecaptionskip\goodbreak
    \addtolength{\splitcaptionheight}{-\abovecaptionskip}
  \fi
  \setbox\splitcaptionbox=\vbox{\interlinepenalty 0
    \reset@font\small{\bfseries#1.} #2\par}
  \vsplit\splitcaptionbox to \splitcaptionheight
  \global\setbox\splitcaptionbox=\box\splitcaptionbox
  \if@cop@home\ifonline\ifnum\csname c@\@captype\endcsname=1 
    \immediate\write\@auxout{\string\gdef\string\@num\@captype{}}
    \hypertarget{\@captype}{}
  \fi\fi\fi}

\newcommand{\splitcaption}[3][\empty]
 {\setlength{\splitcaptionheight}{#3}
  \let\cop@makecaption=\@makecaption
  \let\@makecaption=\@makesplitcaption
  \ifx\empty#1\relax
    \caption{#2}
  \else
    \caption[#1]{#2}
  \fi
  \let\@makecaption=\cop@makecaption}
\makeatother

\newcommand{\mergecaption}{\ifdim\ht\splitcaptionbox>0pt
  \begin{figure}[tp]
  \box\splitcaptionbox
  \end{figure}
\fi}

\begin{document}

\title{Sulphur-bearing molecule survey towards prestellar and starless cores in the Perseus molecular cloud}

\author{J.~Ferrer Asensio, \inst{1} S.~Scibelli,\inst{2,} \thanks{Jansky Fellow of the National Radio Astronomy Observatory.} A.~Pokorny-Yadav,\inst{3,4} L.~Steffes,\inst{5} B.~Kulterer,\inst{4} Y.~Shirley,\inst{5} A.~Meg{\'i}as,\inst{6} I.~Jim{\'e}nez-Serra,\inst{6} A.~Fuente, \inst{6} G.~Esplugues \inst{7}}
\institute{\tiny{\inst{1} RIKEN Cluster for Pioneering Research, Wako-shi, Saitama, 351--0106, Japan\\} \inst{2} National Radio Astronomy Observatory, 520 Edgemont Road, Charlottesville, VA 22903, USA\\ \inst{3} Department of Astronomy, University of California, Berkeley, Berkeley, CA, 94720, USA \\ \inst{4} Department of Astronomy, University of Virginia, 530 McCormick Rd, Charlottesville, Virginia 22904, USA \\ \inst{5} Steward Observatory, University of Arizona, 933 North Cherry Avenue, Tucson, AZ 85721, USA\\     \inst{6} Centro de Astrobiolog\'ia (CAB), CSIC-INTA, Carretera de Ajalvir, km 4, 28850 Torrej\'on de Ardoz, Spain \\   \inst{7} Observatorio Astron\'omico Nacional (OAN), Alfonso XII, 3, 28014 Madrid, Spain}

\titlerunning{Sulphur-bearing molecule survey in the Perseus molecular cloud}
\authorrunning{J. Ferrer Asensio et al.}

\date{Received  ; accepted }

\abstract
{Sulphur is drastically depleted from the gas phase in dense starless and prestellar cores creating the long-standing ``missing sulphur problem''. Observations of sulphur-bearing molecules are crucial to understanding sulphur chemistry, the relationship of cores with their environments, as well as the physical evolution of star-forming regions.}
{We present an observational survey of four sulphur-bearing molecules, HCS$^{+}$, OCS, H$_{2}$CS and SO, towards dense cores in the Perseus molecular cloud with the aim of studying their abundance dependence to core physical properties and environment.}
{The HCS$^{+}$ 1--0, the OCS 3--2 and 4--3, the H$_{2}$CS 1$_{0,1}$--0$_{0,0}$ and the SO 3$_{2}$--2$_{2}$ transitions are observed with the Yebes 40m telescope towards 15 starless and prestellar cores in the Perseus molecular cloud. The lines are fitted with Gaussian profiles and their column densities are calculated assuming the local thermodynamic equilibrium (LTE) approach. Additionally, a comparison with observations of other sulphur-bearing species towards the same core samples presented in a previous study, CS, CCS, and CCCS, is performed.}
{The HCS$^{+}$, OCS, H$_{2}$CS and SO are detected towards 100\%, 93\%, 100\% and 73\% of the source sample, respectively. The derived column densities are (2.2 - 17.0)$\times$10$^{11}$ cm$^{-2}$ for HCS$^{+}$, (1.9 - 25.0)$\times$10$^{12}$ cm$^{-2}$ for OCS, (1.0 - 15.0)$\times$10$^{12}$ cm$^{-2}$ for H$_{2}$CS, and (0.3 - 17.0)$\times$10$^{13}$ cm$^{-2}$ for SO. The molecular column densities found span an order of magnitude or even more in the entire core sample.}
{The abundances of HCS$^{+}$ and SO strongly correlate, inversely and directly, respectively, with the volume density of the cores, reinforcing their early- and late-type molecule status. The correlations found amongst the molecular abundances point at a differentiation between ``carbon chemistry'' and ``oxygen chemistry'' of sulphur-bearing molecules. CS has the strongest abundance correlations with other molecules, as expected from its core role in sulphur chemistry. Cores located in the path of protostellar outflows do not show consistently higher sulphur-bearing molecule abundances compared to the rest of the sample. The chemical age of the cores, as traced by SO/CS, correlates with their physical age, traced using the volume density of the cores, with the exception of Per752, which shows an outlier chemical fingerprint, possibly related to an enhanced cosmic ray ionization rate in the area. Overall, these results provide a crucial observational baseline offering tighter constraints for chemical models of the sulphur chemistry in the earliest stages of star formation.}


\maketitle

\section{Introduction} \label{introduction3}

Sulphur (S) is one of the most abundant elements in the Universe, with an estimated cosmic abundance of S/H $\sim$ 1.32 $\times$ 10$^{-5}$ \citep{yamamoto:17}. However, a major discrepancy exists in astrochemistry known as the ``missing sulphur problem''. Sulphur is seen to remain near its cosmic abundance in diffuse and translucent clouds, while its gas-phase abundance in dense molecular environments drops by a factor of 100 to 1000 \citep{ruffle:99, wakelam:04, fuente:23}. This depletion is particularly relevant in starless and prestellar cores, which mark the beginning of star formation. Starless cores are dense, self-gravitating structures that have yet to form a central protostar. Prestellar cores are a type of starless core that show signs of gravitational contraction, indicating that they are on the verge of stellar birth. These objects are characterised by high densities ($n_{\rm H_2} \sim 10^5$ cm$^{-3}$) and low temperatures ($T$ $\lesssim 10$ K) in their centres, which promote the depletion of molecules from the gas phase onto the surface of dust grains in this region \citep{bergin:07, keto:08}.\ 

Chemical modelling studies suggest the ``missing'' sulphur in prestellar cores is locked into a hidden reservoir, primarily in the surfaces of dust grains (e.g., \citealt{vastel:18} and \citealt{laas:19}). In these dense regions, atomic sulphur is expected to deplete onto grains and undergo successive hydrogenation to form solid H$_{2}$S, which many models identify as the primary sulphur sink \citep{smith:91, vandertak:03, jimenez:11, esplugues:14}. Despite this prediction, solid H$_{2}$S has never been unambiguously detected in ices. Currently, OCS is the only sulphur-bearing molecule firmly identified in interstellar ices, while SO$_{2}$ has only been tentatively detected \citep{mcclure:23, rocha:24, gross:26}. Infrared spectra simulations suggest that detecting sulphur-bearing molecules in ices is still challenging, even with the current most sensitive telescopes like the \textit{James Webb} Space Telescope (JWST) \citep{taillard:25}. This has led researchers to explore alternative reservoirs, such as CH$_{3}$SH, H$_{2}$CS, ammonium salt NH$_{4}$SH, sulphur allotropes trapped in ice mantles, and the recently detected c-C$_{6}$H$_{6}$S \citep{laas:19, cazaux:22, carrascosa:24, vitorino:24, slavicinska:25, araki:26}. \

Observing sulphur-bearing molecules in the gas phase of starless and prestellar cores is necessary to address the ``missing sulphur problem'' and provide critical insights into the chemical and physical evolution of star-forming regions \citep{navarro:20, fuente:23, ghosh:24}. 

Among nearby molecular clouds, the Perseus region serves as an ideal laboratory for studying sulphur chemistry during the early stages of star formation. Located at a distance of approximately 294 pc \citep{zucker:18}, the Perseus molecular cloud hosts a diverse population of starless and prestellar cores embedded within its dense filamentary structures. Previous studies of these cores have revealed a complex chemical inventory, including complex organic molecules (COMs), such as methanol and acetaldehyde, carbon chains, deuterated species, and phosphorous bearing molecules \citep{scibelli:24, scibelli:25a, ferrerasensio:26, pokorny:26}. In addition to these cores, Perseus contains numerous protostellar sources whose feedback may influence the chemistry of neighbouring starless regions \citep{ladd:93, aspin:94, lada:95, hatchell:05, enoch:06, kirk:06, muench:07, gutermuth:08, evans:09, plunkett:13, yang:21}. In this work, we characterise the sulphur-bearing molecular content of a sample of Perseus starless cores to determine how this chemistry depends on local physical and environmental factors.\

This study is structured as follows: the target sulphur-bearing molecules in this study are introduced in Section \ref{obss}, observations are presented in Section \ref{observations}, the description of the methodology used to reduce and fit the data are presented in Section \ref{analysis}, the results are presented in Section \ref{results}, these are interpreted in Section \ref{discussion}, and we summarise the conclusion in Section \ref{conclusions}. Moreover, supplementary information is collected in the Appendix.\

\section{Observed sulphur-bearing species}\label{obss}

HCS$^{+}$, OCS, H$_{2}$CS and SO are some of the most abundant sulphur-bearing molecules detected in the interstellar medium (ISM), playing key roles in its chemistry.\

HCS$^{+}$ is widely detected in the ISM, including dense cores such as Barnard 1, L134N, and TMC-1 \citep{thaddeus:81, irvine:83, turner:96, ruffle:99, fuente:16, vastel:18}. HCS$^{+}$ traces cold, quiescent, extended material \citep{fuente:16}, and it is mainly formed from the hydrogenation of the CS cation, CS$^{+}$ + H$_{2}$ $\rightarrow$ HCS$^{+}$ + H, and also by the reaction S$^{+}$ + CH$_{2}$ $\rightarrow$ HCS$^{+}$ + H in the gas phase \citep{vidal:17, yamamoto:17}.\

OCS has been detected both in the gas phase as well as in ice form on the surfaces of dust grains towards dense cores \citep{matthews:87, palumbo:95, vastel:18, mercimek:22, ghosh:24, miranzo:25}. OCS forms in an efficient way on icy grain mantles, CO + S $\rightarrow$ OCS, and then is released in the gas via non-thermal desorption mechanisms \citep{hasegawa:93, vidal:17, laas:19}. Moreover, gas-phase reaction routes, such as HCS + O $\rightarrow$ H + OCS have also been found to be efficient \citep{laas:19}.\

H$_{2}$CS was first detected in the ISM by \cite{sinclair:73}, and has since been observed towards a great number of starless, prestellar and protostellar sources \citep{marcelino:05, vastel:18, nagy:19, esplugues:22, mercimek:22, spezzano:22}. As a sulphur-bearing analogue of formaldehyde (H$_{2}$CO), H$_{2}$CS can form both through gas-phase reactions and on the surfaces of dust grains, making it sensitive to the dominant chemical pathways operating in cold, dense environments (H$_{3}$CS$^{+}$ + e$^{-}$ on the gas phase, and HCS + H on grain surfaces, \citealt{vidal:17}). H$_{2}$CS is the precursor to more complex sulphur-bearing molecules in the surface of grains such as CH$_{3}$SH, the sulphur analogue of the most abundant COM, methanol (CH$_{3}$OH).\

SO was detected for the first time towards seven galactic sources in \cite{gottlieb:73}. This molecule is known to be a shock tracer arising from shock-induced chemistry or desorption from grain mantles \citep{pineau:93}. It mainly forms through the gas phase reaction OH + S $\rightarrow$ SO + H \citep{laas:19}. The surface formation of SO is considered less efficient due to the high binding energy of the oxygen atom. Nevertheless, a large proportion of SO is thought to be depleted onto dust grains.\

\section{Observations}\label{observations}

\begin{figure*}[htbp]
\centering
\includegraphics[width=\textwidth]{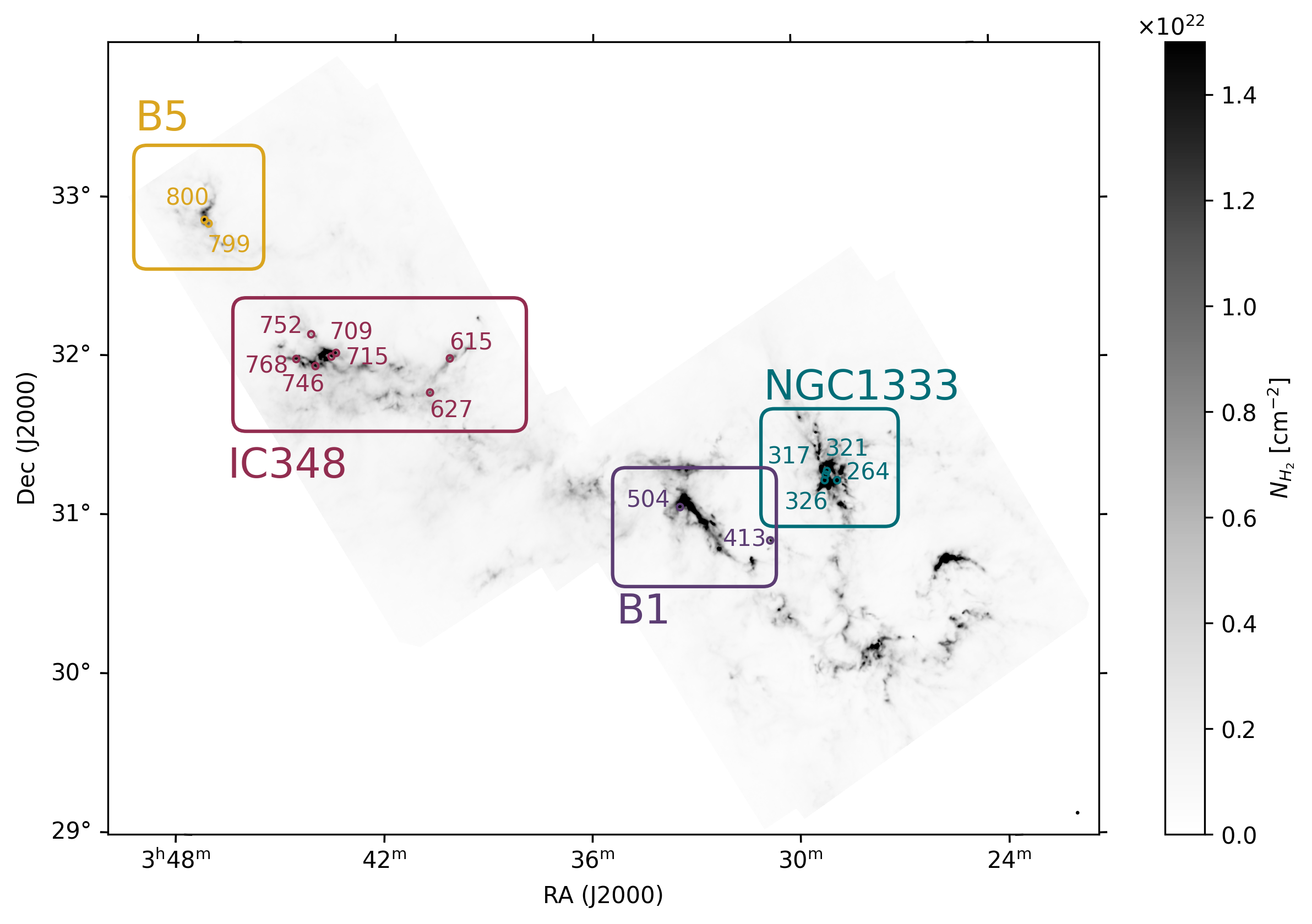}
\caption{Molecular hydrogen column density ($N_{\rm H_2}$) map of the Perseus molecular cloud in grey scale \citep{pezzuto:12,sadavoy:12,sadavoy:14}. The starless and prestellar cores studied in this study are indicated by coloured circles corresponding to the regions within the molecular cloud where they are located. NGC1333 is plotted in teal, B1 in purple, IC348 in red, and B5 in yellow.}
\label{map}
\end{figure*}

The spectral data analysed in this work were observed with the Yebes 40m telescope using the Q-Band receiver, with a bandwidth of 18.5 GHz (31.5 – 50 GHz) \citep{tercero:21}, between 2022 and 2023 towards fifteen starless and prestellar cores in the Perseus molecular cloud (Project ID: 22A022 and 23A025, PI: Scibelli, \citealt{scibelli:24}). These sources were targeted based on the prior detection of acetaldehyde (CH$_{3}$CHO), with the objective of probing these regions for additional COMs. A map of the region and information of the sources studied in the present work can be found in Figure \ref{map} and Table \ref{source}, respectively.\

The spectral resolution and beam sizes for the full bandwidth range are 0.38 -- 0.23 km/s and 36$^{\prime\prime}$ -- 56$^{\prime\prime}$, respectively. The publicly available Python-based scripts\footnote{\url{https://github.com/andresmegias/gildas-class-pipeline/}} developed by \citealt{megias:23}, which calls on the Continuum and Line Analysis Single-dish Software (\textsc{class}), an application from the \textsc{gildas}\footnote{\url{https://www.iram.fr/IRAMFR/GILDAS/}} software \citep{pety:05}, was used for the reduction of the data (see \cite{scibelli:24} for more details).\

The spectral information of the transitions analysed in this work, HCS$^{+}$ 1--0, OCS 3-2 and 4-3, H$_{2}$CS 1$_{0,1}$--0$_{0,0}$ and SO 3$_{2}$--2$_{2}$, is listed in Table \ref{obsdat}.\

\begin{table*}[ht]
\begin{center}
\caption{Starless and prestellar cores in the Perseus molecular cloud targeted in this study. }

\begin{tabular}{ lccccccc } 
\hline\hline
Region & Core $\#$ $^{a}$ & RA & DEC &  Distance$^{b}$ & $N_{\rm H_{2}}$$^{c}$ & $n_{\rm {H_{2}}}$$^{c}$  & $T_{\rm kin}$$^{c}$  \\
 & (\textit{Herschel}) & (J2000) & (J2000) & (pc) & ($\times$ 10$^{22}$ cm$^{-2}$) & ($\times$ 10$^{5}$ cm$^{-3}$) & (K) \\
\hline\hline
\textcolor{NGC1333}{NGC1333} & 264 & 03:28:47.14 & +31:15:11.4 & 294 & 2.06 & 0.75 & 10.8  \\
& 317 & 03:29:04.93 & +31:18:44.4 & 294 & 2.63 & 0.96 & 13.6  \\
& 321 & 03:29:07.17 & +31:17:22.1 & 294 & 3.65 & 1.33 & 12.6  \\
& 326 & 03:29:08.97 & +31:15:17.2 & 294 & 7.88 & 2.87 & 12.3  \\
\hline
\textcolor{B1}{B1} & 413 & 03:30:46.74 & +30:52:44.8 & 297 & 1.17 & 0.42 & 10.5  \\
& 504 & 03:33:25.31 & +31:05:37.5 & 297 &1.80 & 0.65 & 9.7  \\
\hline
\textcolor{IC348}{IC348} & 615 & 03:40:14.92 & +32:01:40.8 & 314 & 1.03 & 0.35 & 10.3  \\
& 627 & 03:40:49.53 & +31:48:40.5 & 314 & 0.93 & 0.32 & 12.4  \\
& 709 & 03:43:38.06 & +32:03:07.4 & 314 & 1.48 & 0.51 & 13.5  \\
& 715 & 03:43:46.34 & +32:01:43.5 & 314 & 1.45 & 0.49 & 14.8  \\
& 746 & 03:44:14.38 & +31:58:00.7 & 314 & 1.42 & 0.49 & 10.6  \\
& 752 & 03:44:23.10 & +32:10:01.0 & 314 & 0.63 & 0.22 & 14.7  \\
& 768 & 03:44:48.83 & +32:00:31.6 & 314 & 1.43 & 0.49 & 10.8  \\
\hline
\textcolor{B5}{B5} & 799 & 03:47:31.31 & +32:50:56.9 & 325 & 1.01 & 0.33 & 10.4  \\
& 800 & 03:47:38.97 & +32:52:16.6 & 325 & 1.93 & 0.64 & 11.7  \\
\hline
\label{source}
\end{tabular}

\tablefoot{$^{a}$ The labelling of the cores correspond to the one defined in \cite{pezzuto:21}. $^{b}$ The distances of the cores are derived in \cite{pezzuto:21}. $^{c}$ The $N_{\rm H_{2}}$, $n_{\rm {H_{2}}}$, and $T_{\rm kin}$ values are taken from \cite{scibelli:24}, see Section \ref{analysis} for more details.}
\end{center}
\end{table*}

\begin{table*}[ht]
\begin{center}
\caption{Observed molecular transitions.}

\begin{tabular}{ lcccc } 
\hline\hline
Molecule & Transition &  Frequency &  $E_\mathrm{up}$ & $A_{ij}$  \\
 & &  (MHz) & (K) & ($\times$ 10$^{-7}$ s$^{-1}$)  \\
\hline\hline
HCS$^{+}$ & 1--0 & 42674.1954 & 2.05 & 11.56   \\
OCS & 3--2 & 36488.8121 & 3.50 & 1.24   \\
& 4--3 & 48651.6043 & 5.84 & 3.05   \\
H$_{2}$CS & 1$_{0,1}$--0$_{0,0}$ & 34351.4300 & 1.65 & 4.28   \\
SO & 3$_{2}$--2$_{2}$ & 36202.0220 & 21.05 & 1.94   \\
\hline
\label{obsdat}
\end{tabular}

\tablefoot{ The spectral information for the molecular transitions including the frequency, the upper energy ($E_{\rm up}$), and the Einstein coefficient ($A_{ij}$), was taken from The Cologne Database for Molecular Spectroscopy (CDMS)\footnote{\url{https://cdms.astro.uni-koeln.de}} (HCS$^{+}$: \citealt{margules:03, botschwina:85}, OCS: \citealt{winton:70, wang:73, kukolich:74, dubrulle:80, golubiatnikov:05}, H$_{2}$CS: \citealt{muller:19, maeda:08, clouthier:94, johnson:71, johnson:70, fabricant:77, mcnaughton:93}, SO: \citealt{amano:74, clark:76, tiemann:82, lovas:92, cazzoli:94, klaus:96, bogey:97, powell:64}.)}
\end{center}
\end{table*}

\section{Analysis} \label{analysis}

The fit of the observed spectra was carried out with the Python ``pyspeckit'' package \citep{ginsburg:11, ginsburg:22}. The results of this fit can be found in Appendix \ref{obstran}. The fit of the SO 3$_{2}$--2$_{2}$ transition towards the core Per627 with ``pyspeckit'' failed resulting in unusually large uncertainties for the local standard of rest velocity ($v_{\rm LSR}$) and the full width half maximum (FWHM) of several orders of magnitude. We attribute this uncertainty to the line being comprised effectively by only two spectral channels resulting from the limited spectral resolution of the observations at the frequency of the transition with a value of $\sim$ 0.32 km/s. For this transition the area under the line is integrated using a Python routine, and by assuming a fixed line width of 0.5 km/s, which is consistent with the FWHM of other molecules detected towards Per627, its peak intensity is derived. This transition is marked with an asterisk ($*$) in Table \ref{obstrans} in the Appendix. The transitions are considered detected if the signal to noise ratio in peak intensity (S/N) is $\geq$3. Although most molecules have only a single transition detected above 3$\sigma$ per core, mostly due to the fact that these are the only strong transitions of these molecules in the studied bandwidth, the detection claims are considered reliable because HCS$^{+}$, OCS, H$_{2}$CS and SO are known to be ubiquitous throughout the ISM. Furthermore, no other plausible molecular line candidates exist near the observed frequencies.  \

The column density is computed assuming local thermodynamic equilibrium (LTE). First, the optical depth ($\tau$) of the transitions is calculated for a range of excitation temperatures ($T_{\rm {ex}}$), 5 - 30 K, to assess which LTE approximation, whether the optically thin or optically thick one, is to be used to compute the column densities of the observed molecules. For this purpose, the following equation taken from \cite{mangum:15} is used:

\begin{equation}
\tau = -\ln \left[ 1 - 
\frac{T_{\mathrm{MB}}}
{J_\nu(T_{\mathrm{ex}}) - J_\nu(T_{\mathrm{bg}})}
\right].
\label{tau}
\end{equation}

In Equation \ref{tau} the $T_{\rm MB}$ is the main beam temperature of the line and $J_\nu(T_{i})$, being $T_{i}$ = $T_{\rm ex}$ or $T_{\rm bg}$, where $T_{\rm bg}$ is the background cosmic microwave background (CMB) temperature, is:

\begin{equation}
J_\nu(T_{i}) = \frac{h\nu}{k} \left[ \exp\!\left(\frac{h\nu}{kT_{i}}\right) - 1 \right]^{-1},
\end{equation}
where $h$ is the Planck constant, $\nu$ is the frequency and $k$ is the Boltzmann constant.
The analysed transitions are always optically thin with $\tau$ $<$ 0.2. Consequently, the column density of the molecules is calculated with the optically thin LTE approximation equation:
\begin{equation}
N_{\rm tot} =
\frac{8\pi k \nu^2}{h c^3 A_{\rm ul}}
\frac{Q(T_{\mathrm{ex}})}{g_{\rm up}}
\exp\!\left(\frac{E_{\rm up}}{T_{\mathrm{ex}}}\right)
\frac{I}{J_\nu(T_{\mathrm{ex}}) - J_\nu(T_{\mathrm{bg}})},
\end{equation}
from \cite{mangum:15}, where $I$ is the integrated intensity of the line, $c$ is the light speed, $Q(T_{\rm ex})$ is the partition function at $T_{\rm ex}$, $g_{up}$ is the upper level degeneracy factor and $E_{up}$ is the energy of the upper level. For the LTE approach a $T_{\rm {ex}}$ needs to be assumed. In this study the column densities are calculated assuming a $T_{\rm {ex}}$ of 10 K, which is a common practice to derive column densities towards starless and prestellar cores \citep{jimenez:16, jimenez:21, scibelli:24}. The column density uncertainties include both observational and systematic contributions. The observational component is obtained by propagating the rms noise, adopted as the uncertainty on $T_{\rm {MB}}$, as well as the uncertainty on the FWHM. The systematic uncertainty associated with the excitation temperature is estimated by recalculating the column density assuming $T_{\rm {ex}}$ = 5 and 15 K. The resulting deviations from the nominal value at $T_{\rm {ex}}$ = 10 K are combined in quadrature with the observational uncertainty to derive the final error bars. The range of $T_{\rm {ex}}$ considered for the uncertainty propagation is guided by the values of OCS, H$_{2}$CS and SO found towards L1544: 7.5$\pm$0.6, 12.3$\pm$0.7 and 9.7$\pm$1.8 K, respectively, obtained from assuming LTE conditions with an Markov Chain Monte Carlo (MCMC) approach in \cite{vastel:18}.\

The column densities are also derived with the non-LTE RADEX approach in Appendix \ref{columndensity}. Nevertheless, due to limited constraints in the RADEX fits, and the fact that generally the LTE and non-LTE column densities agree or deviate by less than a factor 2 within uncertainties, from this point on the column densities derived with LTE are discussed. A more in depth description of the non-LTE approach, as well as comparison of the two methods, can be found in Appendix \ref{columndensity}.

With the aim of exploring correlations between the physical parameters of the cores and the molecular abundances, as well as between the molecular abundances themselves, both the Pearson and the Spearman rank correlation coefficients, $r$ and $\rho$, respectively, are used. The Pearson coefficient measures the strength and direction of a linear relationship between two variables. In contrast, the Spearman rank coefficient assesses the degree of monotonic association, relying only on the ranks of the data rather than their actual values, making it less sensitive to outliers or non-linear trends.\

\section{Results} \label{results}

The resulting parameters of Gaussian fitting ($T_{\rm MB}$, $v_{\rm LSR}$ and FWHM) of the observed HCS$^{+}$, OCS, H$_{2}$CS and SO transitions towards all of the core sample can be found in Appendix \ref{obstran}. The S/N of these transitions range between 3 and 108, with a median of 13. The Gaussian fits themselves are presented in Figure \ref{grid1}. The detection statistics of these molecules towards the source sample can be seen in Figure \ref{detstat}. To ease the navigation of this paper the results related to HCS$^{+}$, OCS, H$_{2}$CS and SO are presented in green, yellow, orange and red, respectively.\

In this section the column densities calculated with the LTE approach are presented (Table \ref{coldenst}).  \

The results for HCS$^{+}$, OCS, H$_{2}$CS and SO are presented in Sections \ref{hcs+}, \ref{ocs}, \ref{h2cs} and \ref{so}. The molecular abundances are plotted against $n_{\rm H_2}$ and $T_{\rm kin}$, respectively, in Section \ref{physs}. Lastly, the molecular abundances are compared to each other in Section \ref{rat}.\

\begin{figure*}[htbp]
\centering
\includegraphics[width=15cm]{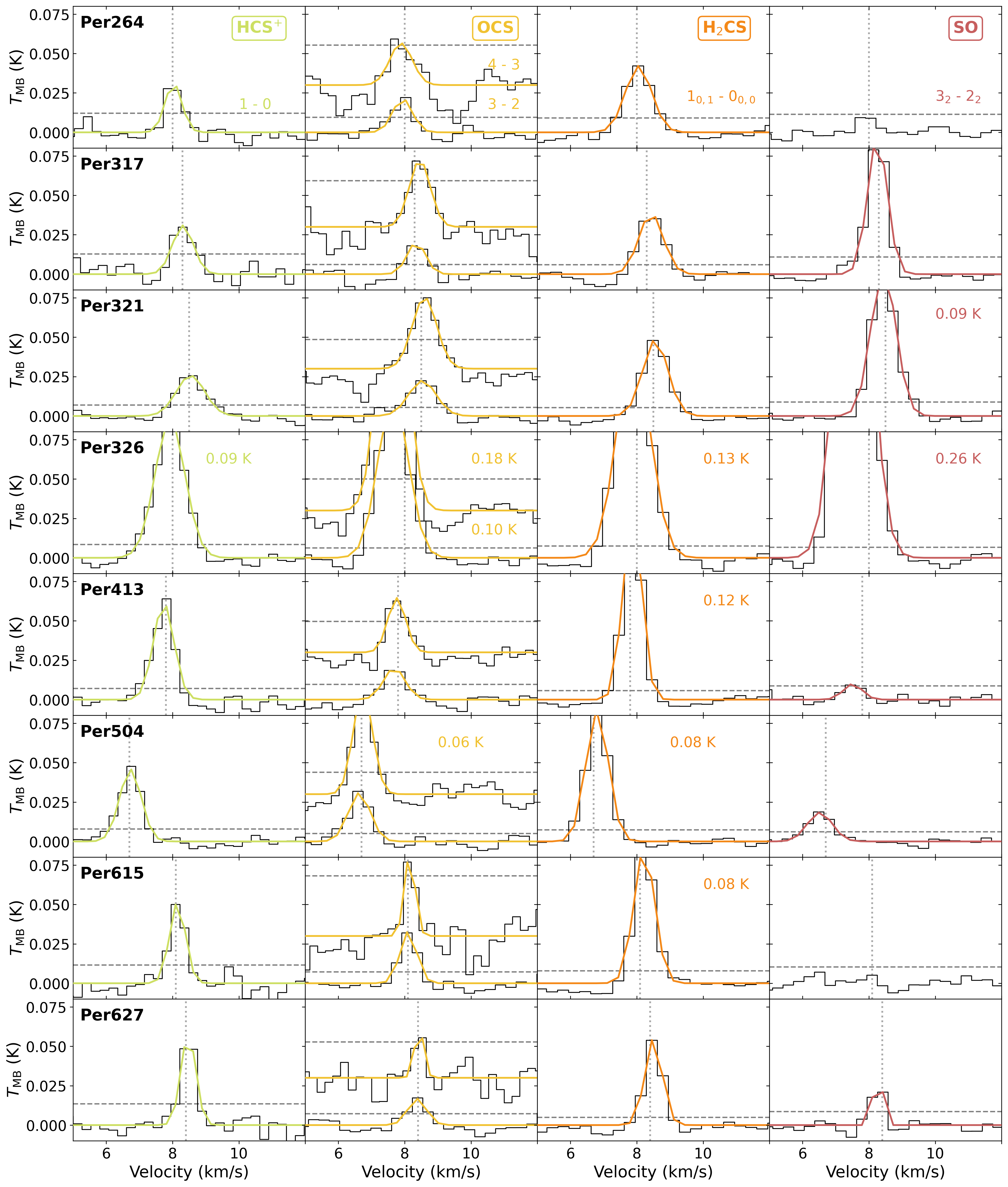}
\caption{Spectra of HCS$^{+}$, OCS, H$_{2}$CS and SO, in the first, second, third and fourth columns, respectively, observed for each core of the sample, in each row (black). The two transitions of OCS for the same core are shown in the same subplot with a vertical offset of 30 mK for clarity. The Gaussian fits from ``pyspeckit'' are overplotted in green, yellow, orange and red for HCS$^{+}$, OCS, H$_{2}$CS and SO, respectively. All of the plots share $x$ and $y$ axes for comparison. And therefore some of the transition peaks are cut off. Spectra without a Gaussian fit indicate non-detection. Horizontal gray dashed lines indicate 3$\sigma$ levels. Vertical gray dotted lines indicate typical $v_{\rm LSR}$ for each of the cores, derived from the lines fitted in this work and in \cite{ferrerasensio:26}.}
\label{grid1}
\end{figure*}

\begin{figure*}[htbp]
\ContinuedFloat 
\centering
\includegraphics[width=15cm]{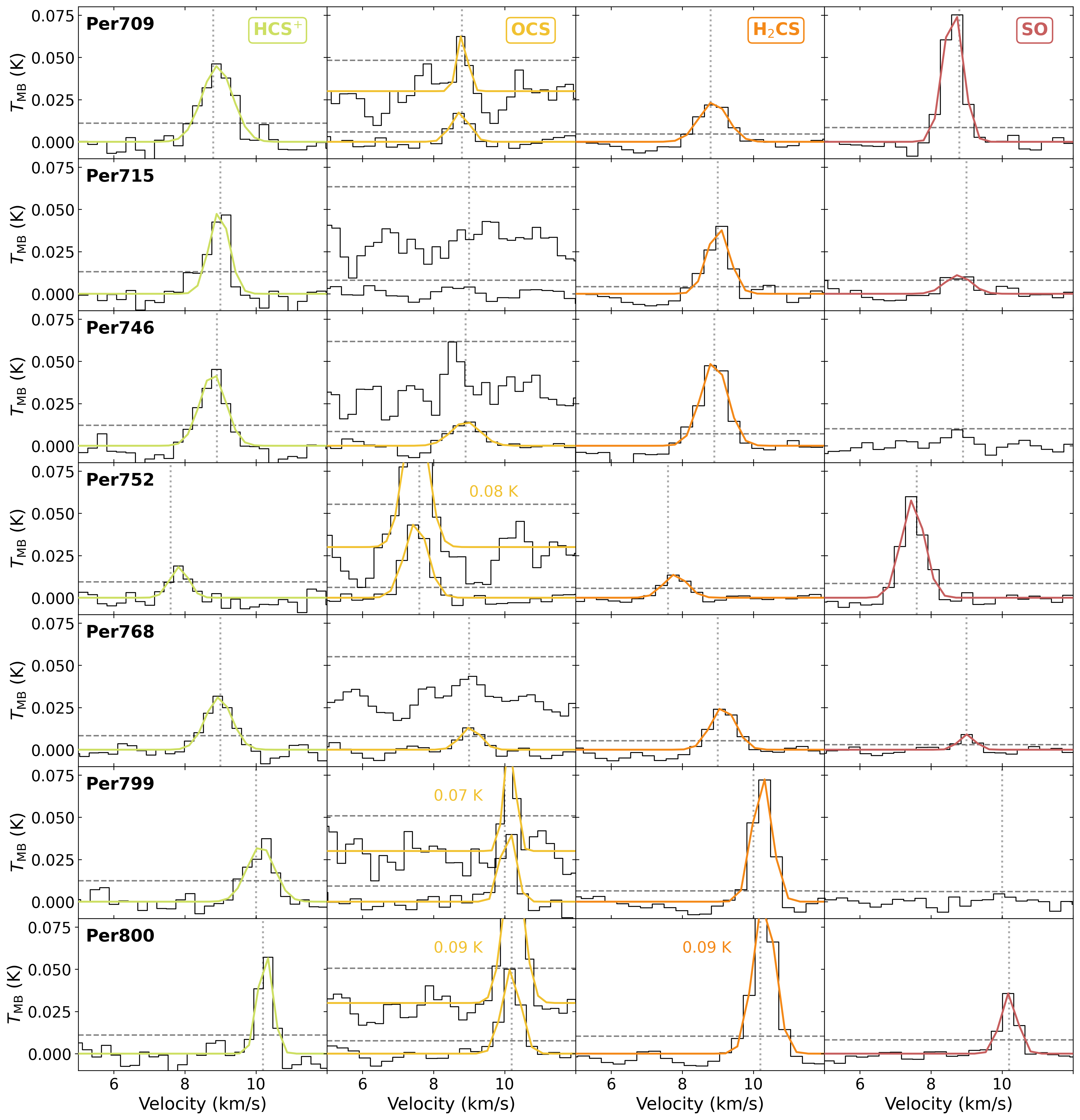}
\caption{Continued.}
\label{grid2}
\end{figure*}

\begin{figure}[h]
\centering
\includegraphics[width=6cm]{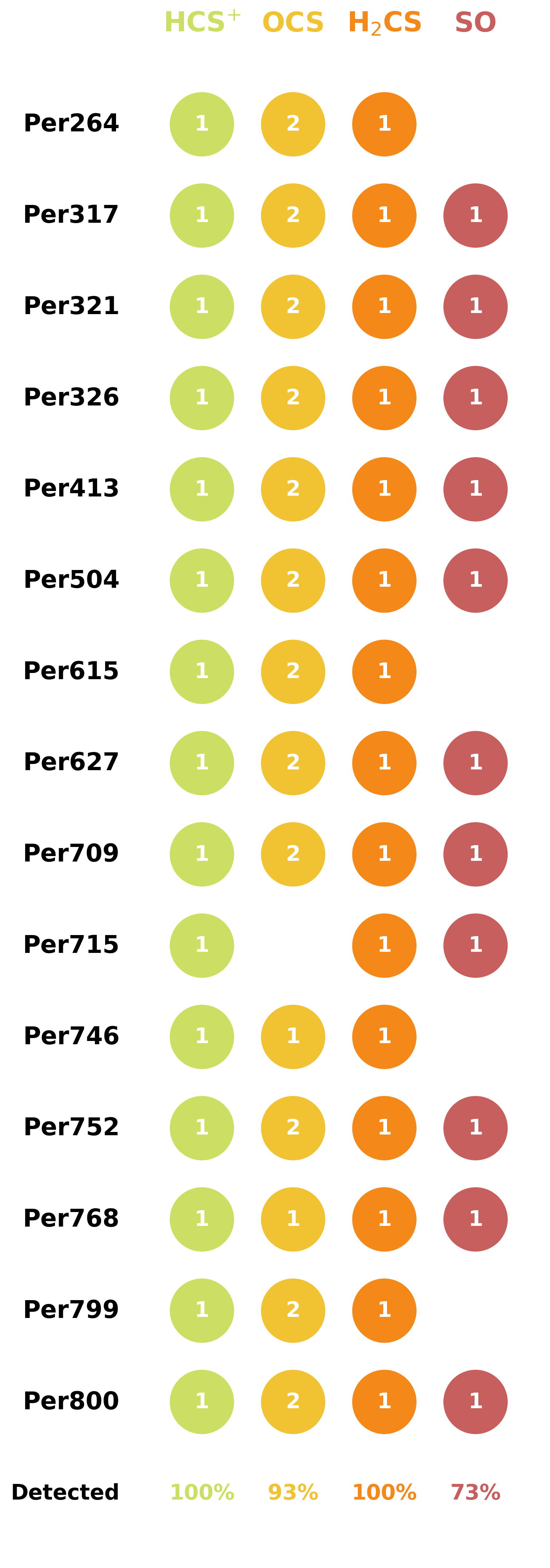}
\caption{Detection statistics of HCS$^{+}$, OCS, H$_{2}$CS and SO for each core in green, yellow, orange and red, respectively. The presence of a circle indicates that the molecule has been detected towards the core. The number within the circle corresponds to the number of transitions detected. The last row shows the percentage of cores the molecule in that specific column has been detected towards. }
\label{detstat}
\end{figure}

\subsection{HCS$^{+}$} \label{hcs+}

HCS$^{+}$ is detected towards the totality of the core sample (Figure \ref{detstat}). The column density calculated with the LTE approach, as described in Section \ref{analysis}, gives a range of (2.2 - 17.0)$\times$10$^{11}$ cm$^{-2}$ (Table \ref{coldenst}) with a median of (5.6 $\pm$ 0.5) $\times10^{11}$. The core with the highest HCS$^{+}$ column density is Per326 and the lowest is Per752.\

\subsection{OCS} \label{ocs}
OCS is detected towards 14 cores of the total sample of 15 (93\%). The core where OCS is not detected is Per715. The two 3--2 and 4--3 transitions were detected towards all of the cores except Per746 and Per768, where only the 3--2 transition was detected (Figure \ref{detstat}). The column density calculated with the LTE approach for OCS has a range of (1.9 - 25.0)$\times$10$^{12}$ cm$^{-2}$ with a median of (3.7 $\pm$ 0.4) $\times10^{12}$ for all the cores. The core with the largest OCS column density is Per326 and the one with the lowest is Per627.\

\subsection{H$_{2}$CS} \label{h2cs}

H$_{2}$CS is detected towards the entire core sample (Figure \ref{detstat}). The column density calculated with the LTE approximation has a range of (1.0 - 15.0)$\times$10$^{12}$ cm$^{-2}$ with a median of (4.0 $\pm$ 0.4) $\times10^{12}$ for all the cores. The core with the largest H$_{2}$CS column density is Per326 and the core with the smallest column density is Per752.\

\subsection{SO} \label{so}

SO is detected in 11 out of the 15 target sources (73\%). The cores where SO is not detected are Per264, Per615, Per746 and Per799. The column density range, excluding the nondetections is (0.3 - 17.0)$\times$10$^{13}$ cm$^{-2}$ with a median of (1.1 $\pm$ 0.9) $\times10^{13}$ cm$^{-2}$. The core with the largest SO column density is Per326 and the one with the lowest is Per768.\

\subsection{Molecular abundance correlation with $n_{\rm H_2}$ and $T_{\rm kin}$} \label{physs}

\begin{figure}[h]
\centering
\includegraphics[width=8cm]{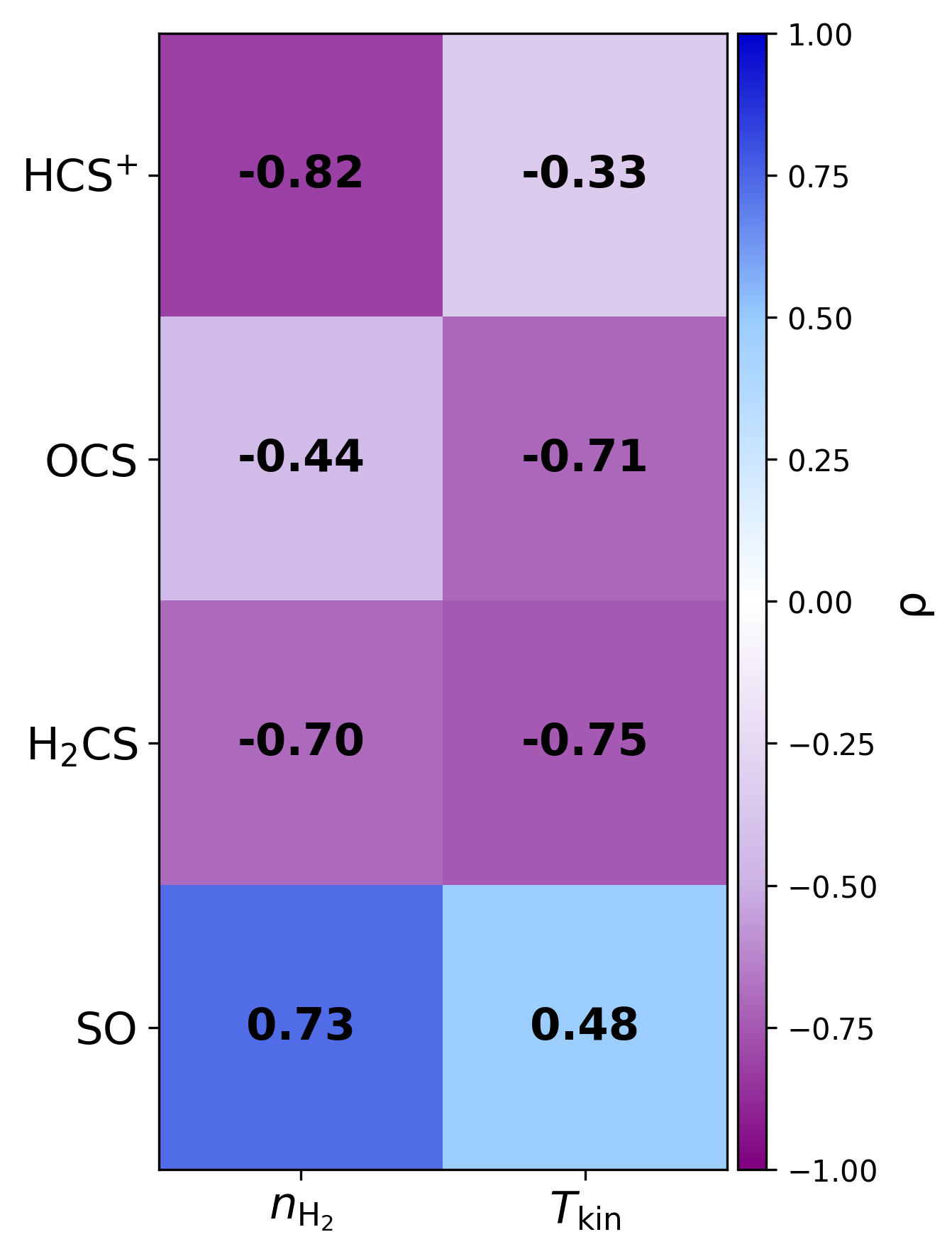}
\caption{Spearman correlation coefficients ($\rho$) of molecular abundances vs $n_{\rm H_2}$ and $T_{\rm kin}$. For all correlations cores Per326 and Per752 are excluded. The squares plotted in blue correspond to $\rho >$ 0 and the squares plotted in purple correspond to $\rho <$ 0.}
\label{physgrid}
\end{figure}

The molecular abundances $X_{\rm mol}$ = ($N_{\rm mol}$/$N_{\rm H_{2}}$) are calculated using the molecular hydrogen column densities in Table \ref{source} and are plotted against the volume density ($n_{\rm H_2}$) and kinetic temperature ($T_{\rm kin}$) of the cores (Figure \ref{phys}). The average and median abundances taking into account all the detections for HCS$^{+}$, OCS, H$_{2}$CS and SO are summarized in Table \ref{abunt}. The difference between the average and median values for some molecules is due to the presence of outlier molecular abundances. The abundances for all molecules towards Per326 have been excluded from the analysis due to their outlier values, as seen from the comparison of the Pearson ($r$) and Spearman rank ($\rho$) coefficients (see Appendix \ref{physa}). Moreover, the abundances of OCS and SO towards Per752 have also been classified as outliers, determined with the same process as the abundances towards Per326, and excluded towards all the sample for consistency. To properly treat the correlations which may include some points showing mild outlier behaviour, the Spearman rank coefficient method is chosen to analyse the trends. The $\rho$ coefficients are summarised in Figure \ref{physgrid}. The correlation plots, which also show the $\rho$ values including the complete core sample, are shown in full in Figure \ref{phys}. The $\rho$ is separated into different tiers depending on its value: from 0.00 to 0.19 the correlation is deemed as negligible, from 0.20 to 0.39 as a weak correlation, from 0.40 to 0.59 as a moderate correlation, from 0.60 to 0.79 as a strong correlation, and finally, from 0.80 to 1.00 as a very strong correlation. The same applies for the negative $\rho$ values. Taking into account this definition the abundance of HCS$^{+}$ has a very strong inverse correlation ($\rho$ = -0.82) with $n_{\rm H_2}$, and SO has a strong direct correlation ($\rho$ = 0.73) with $n_{\rm H_2}$. The OCS abundance has a moderate inverse correlation ($\rho$ = -0.44), while the H$_{2}$CS abundance has a strong inverse correlation ($\rho$ = -0.70) with $n_{\rm H_2}$. Removing Per752 strengthens both the correlation between HCS$^{+}$ and H$_{2}$CS with $n_{\rm H_2}$, compared to the correlations including Per752, $\rho$ = -0.66 and -0.53, respectively. For HCS$^{+}$, the correlation with $n_{\rm H_2}$ is always at least strong. On the other hand, removing Per752 from the H$_{2}$CS vs $n_{\rm H_2}$ comparison changes the correlation from moderate to strong, and thus the correlation shown in Figure \ref{physgrid} is less robust as it depends substantially on Per752. The molecular abundance correlations with $T_{\rm kin}$ show a strong inverse correlation for OCS and H$_{2}$CS ($\rho$ = -0.71 and -0.75, respectively), a weak inverse correlation with the HCS$^{+}$ abundance ($\rho$ = -0.33) and a moderate direct correlation with SO ($\rho$ = 0.48). It needs to be noted that removing Per752 for HCS$^{+}$ and H$_{2}$CS vs $T_{\rm kin}$ does not change the results significantly. The HCS$^{+}$, OCS and H$_{2}$CS abundances tend to show inverse correlation tendencies with both $n_{\rm H_2}$ and $T_{\rm kin}$, contrarily to SO which shows direct correlations. In addition, HCS$^{+}$ and SO show stronger correlations with $n_{\rm H_2}$, while OCS and H$_{2}$CS show stronger correlations with $T_{\rm kin}$.  \

\begin{table*}[ht]
\begin{center}
\caption{Column densities calculated with the local-thermodynamic-equilibrium (LTE) method in units of cm$^{-2}$. }

\resizebox{15cm}{!}{
\begin{tabular}{lcccc}
\hline
Core number & HCS$^+$ & OCS & H$_2$CS & SO \\
\hline
264 & (3.1 $\pm$ 0.7)$\times$10$^{11}$ & (3.5 $\pm$ 0.8)$\times$10$^{12}$ & (3.9 $\pm$ 0.7)$\times$10$^{12}$ & $<$3.6$\times$10$^{12}$ \\
317 & (4.1 $\pm$ 1.0)$\times$10$^{11}$ & (3.7 $\pm$ 0.9)$\times$10$^{12}$ & (3.0 $\pm$ 0.5)$\times$10$^{12}$ & (3.0 $\pm$ 1.5)$\times$10$^{13}$ \\
321 & (4.8 $\pm$ 1.0)$\times$10$^{11}$ & (5.4 $\pm$ 1.1)$\times$10$^{12}$ & (4.2 $\pm$ 0.7)$\times$10$^{12}$ & (4.6 $\pm$ 2.3)$\times$10$^{13}$ \\
326 & (1.7 $\pm$ 0.3)$\times$10$^{12}$ & (2.5 $\pm$ 0.4)$\times$10$^{13}$ & (1.5 $\pm$ 0.2)$\times$10$^{13}$ & (1.7 $\pm$ 0.8)$\times$10$^{14}$ \\
413 & (8.8 $\pm$ 1.6)$\times$10$^{11}$ & (3.1 $\pm$ 0.7)$\times$10$^{12}$ & (7.6 $\pm$ 1.2)$\times$10$^{12}$ & (3.4 $\pm$ 2.0)$\times$10$^{12}$ \\
504 & (6.9 $\pm$ 1.3)$\times$10$^{11}$ & (6.0 $\pm$ 1.1)$\times$10$^{12}$ & (6.6 $\pm$ 1.1)$\times$10$^{12}$ & (9.1 $\pm$ 4.7)$\times$10$^{12}$ \\
615 & (5.3 $\pm$ 1.1)$\times$10$^{11}$ & (3.4 $\pm$ 0.7)$\times$10$^{12}$ & (5.9 $\pm$ 0.9)$\times$10$^{12}$ & $<$4.2$\times$10$^{12}$ \\
627 & (4.8 $\pm$ 1.0)$\times$10$^{11}$ & (1.9 $\pm$ 0.4)$\times$10$^{12}$ & (3.3 $\pm$ 0.5)$\times$10$^{12}$ & (3.0 $\pm$ 1.6)$\times$10$^{12}$ \\
709 & (8.8 $\pm$ 1.8)$\times$10$^{11}$ & (2.0 $\pm$ 0.4)$\times$10$^{12}$ & (2.0 $\pm$ 0.4)$\times$10$^{12}$ & (2.7 $\pm$ 1.3)$\times$10$^{13}$ \\
715 & (5.8 $\pm$ 1.2)$\times$10$^{11}$ & $<$1.5$\times10^{11}$ & (2.9 $\pm$ 0.5)$\times$10$^{12}$ & (4.8 $\pm$ 2.6)$\times$10$^{12}$ \\
746 & (7.3 $\pm$ 1.5)$\times$10$^{11}$ & (2.5 $\pm$ 0.6)$\times$10$^{12}$ & (4.1 $\pm$ 0.7)$\times$10$^{12}$ & $<$3.5$\times$10$^{12}$ \\
752 & (2.2 $\pm$ 0.5)$\times$10$^{11}$ & (8.5 $\pm$ 1.6)$\times$10$^{12}$ & (1.0 $\pm$ 0.2)$\times$10$^{12}$ & (2.5 $\pm$ 1.3)$\times$10$^{13}$ \\
768 & (5.2 $\pm$ 1.1)$\times$10$^{11}$ & (2.4 $\pm$ 0.7)$\times$10$^{12}$ & (2.0 $\pm$ 0.3)$\times$10$^{12}$ & (2.7 $\pm$ 1.4)$\times$10$^{12}$ \\
799 & (6.3 $\pm$ 1.4)$\times$10$^{11}$ & (4.0 $\pm$ 0.7)$\times$10$^{12}$ & (4.6 $\pm$ 0.7)$\times$10$^{12}$ & $<$2.3$\times$10$^{12}$ \\
800 & (5.2 $\pm$ 1.0)$\times$10$^{11}$ & (7.7 $\pm$ 1.5)$\times$10$^{12}$ & (6.3 $\pm$ 1.0)$\times$10$^{12}$ & (1.1 $\pm$ 0.6)$\times$10$^{13}$ \\
\hline
\textbf{Median} & (5.6 $\pm$ 0.5)$\times$10$^{11}$ & (3.7 $\pm$ 0.4)$\times$10$^{12}$ & (4.0 $\pm$ 0.4)$\times$10$^{12}$ & (1.1 $\pm$ 0.9)$\times$10$^{13}$ \\
\hline
\end{tabular}
}
\tablefoot{The column densities are calculated assuming a $T_{\rm ex}$ = 10 $\pm$ 5 K. The upper limits correspond to the 3$\sigma$ level. The last row shows the median molecular column density for all the cores in the sample with detections. The uncertainty of the median corresponds to the 16th and 84th percentiles, equivalent to a 1-sigma confidence interval, derived via Monte Carlo simulations.}
\label{coldenst}
\end{center}
\end{table*}

\begin{table}[ht]
\begin{center}
\caption{Average and median molecular abundances across the core sample excluding non-detections. }

\resizebox{8cm}{!}{%
\begin{tabular}{lcccc}
\hline
Core number & HCS$^+$ & OCS & H$_2$CS & SO \\
\hline
\textbf{Average} & 4.0$\times$10$^{-11}$ & 3.2$\times$10$^{-10}$ & 2.8$\times$10$^{-10}$ & 1.1$\times$10$^{-9}$ \\
\textbf{Median} & 3.8$\times$10$^{-11}$ & 1.9$\times$10$^{-10}$ & 2.0$\times$10$^{-10}$ & 0.6$\times$10$^{-9}$ \\
\hline
\end{tabular}
}
\label{abunt}
\end{center}
\end{table}

\subsection{Molecular abundance ratios} \label{rat}

The abundances of the different molecules are also plotted against each other with the aim of learning about the dependence between them. In this section we present the correlation of the abundances of HCS$^{+}$, OCS, H$_{2}$CS, SO, CS, CCS and CCCS. The column densities of the CS, CCS and CCCS molecules towards the same sources in Perseus are taken from \cite{pokorny:26}.\ 

\begin{figure}[h]
\centering
\includegraphics[width=8cm]{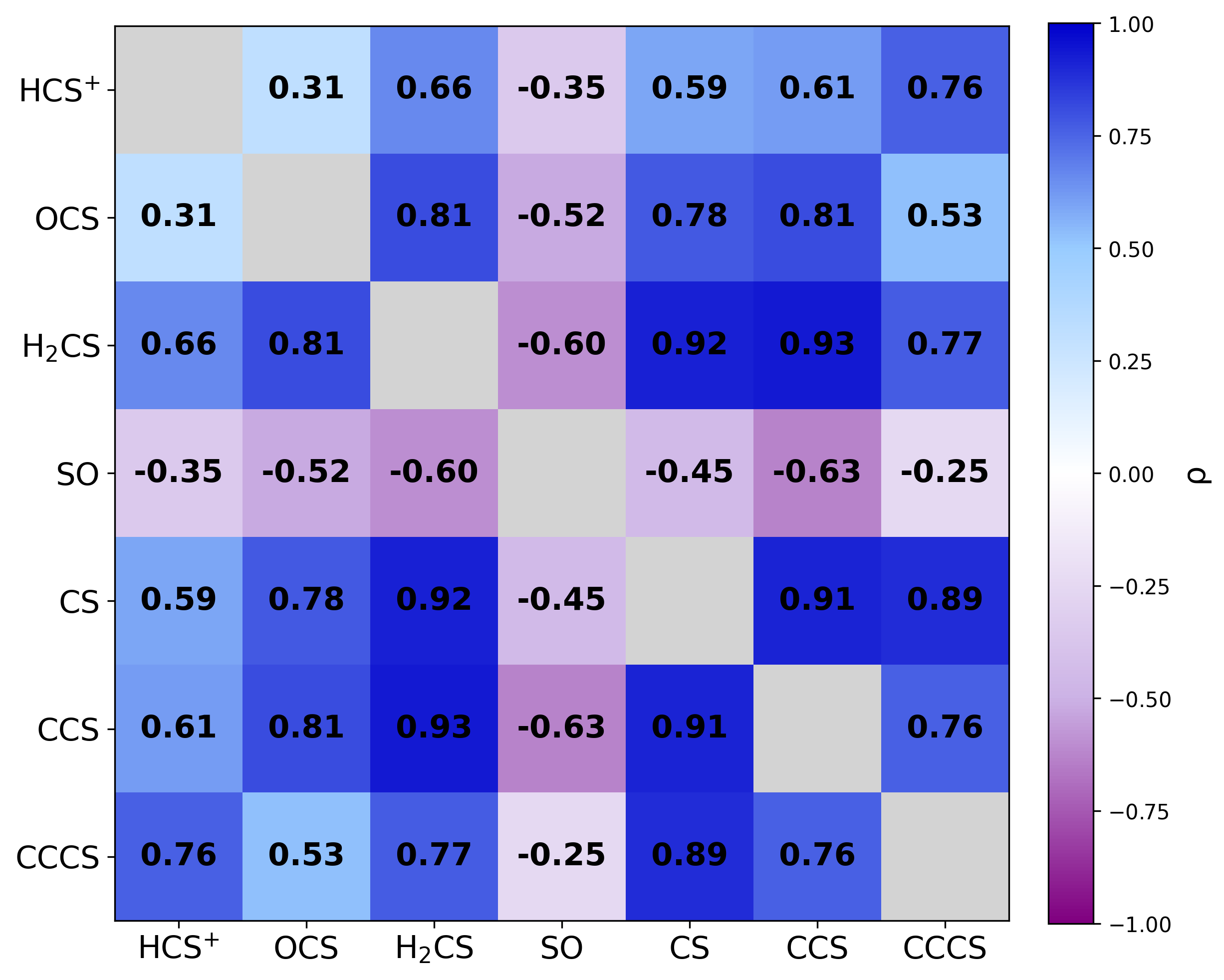}
\caption{Spearman correlation coefficients of pairs of molecular abundances for the core sample. Cores Per326 and Per752 are excluded from all correlations. The squares plotted in blue correspond to $\rho >$ 0 and the squares plotted in purple correspond to $\rho <$ 0. The diagonal, which corresponds to the correlation of a molecule with itself, is plotted in gray.}
\label{moleculegrid}
\end{figure}

The core Per326 was classified as an outlier on the abundances against $n_{\rm H_2}$ and $T_{\rm kin}$ plots mostly for its great influence on the $n_{\rm H_2}$ correlation, due to its large volume density value. When comparing molecular abundances Per326 has less of an influence, but for consistency reasons is also excluded. In this case, the abundances of both OCS and SO towards the Per752 core are seen to behave as outliers and are therefore excluded from all the comparisons for consistency. Including these outlier abundances alters some of the $\rho$ values by up to a factor of $\sim$2, significantly weakening correlations of $n_{\rm H_2}$ and $T_{\rm kin}$ with some molecular abundances. For more information refer to Appendix \ref{molcompa}. The $\rho$ coefficient of each of the molecule pairs studied are plotted in Figure \ref{moleculegrid}. The $\rho$ values including the entire core sample, as well as the correlation plots themselves, can be found in Figure \ref{triangle}.  \

The same correlation-strength tiers as defined in Section \ref{physs} are used to describe the results. The pair of molecular abundances that show very strong correlations are H$_{2}$CS vs CCS ($\rho$ = 0.93), H$_{2}$CS vs CS ($\rho$ = 0.92), CS vs CCS ($\rho$ = 0.91), CS vs CCCS ($\rho$ = 0.89), H$_{2}$CS vs OCS ($\rho$ = 0.81) and OCS vs CCS ($\rho$ = 0.81). The ones that show strong correlations are OCS vs CS ($\rho$ = 0.78), H$_{2}$CS vs CCCS ($\rho$ = 0.77), HCS$^{+}$ vs CCCS ($\rho$ = 0.76), CCS vs CCCS ($\rho$ = 0.76), HCS$^{+}$ vs H$_{2}$CS ($\rho$ = 0.66), SO vs CCS ($\rho$ = -0.63), HCS$^{+}$ vs CCS ($\rho$ = 0.61) and H$_{2}$CS vs SO ($\rho$ = -0.60). The ones with mild correlations are HCS$^{+}$ vs CS ($\rho$ = 0.59), OCS vs CCCS ($\rho$ = 0.53) and OCS vs SO ($\rho$ = -0.52). Finally, the ones with weak correlations are SO vs CS ($\rho$ = -0.45), HCS$^{+}$ vs SO ($\rho$ = -0.35), HCS$^{+}$ vs OCS ($\rho$ = 0.31) and SO vs CCCS ($\rho$ = -0.25). The molecule with the strongest correlations with other molecules is CS, and the one with the weakest is SO. Moreover, SO has abundance inverse correlations with all the molecules, while the rest of abundance comparisons show direct correlations.\

\section{Discussion} \label{discussion}

The HCS$^{+}$ and H$_{2}$CS molecules have been detected towards the entirety of the sample. On the other hand, OCS has been detected towards all the cores except for Per715, and SO has been detected towards all the cores except for Per264, Per615, Per746 and Per799. There is no overlap with the non-detected OCS and SO cores. The core with the highest column densities across all molecules is Per326, while the core with the lowest column density varies. Thus, the non-detections of OCS and SO towards these cores do not seem to be related to a general sulphur-bearing molecule low abundance, but to other variables. The only core with the OCS non-detection, Per715, is the core with the highest $T_{\rm kin}$. As seen in Section \ref{physs} the abundance of OCS has a strong inverse correlation with $T_{\rm kin}$. Core Per752, with the same $T_{\rm kin}$ as Per715, has a high OCS abundance (Figure \ref{phys}), but in this study it is categorised as an outlier (see Section \ref{physs}). Therefore, while temperature may contribute to the low OCS abundance in Per715, it is unlikely to be the only factor, and additional processes, e.g. local physical or chemical conditions, are likely involved. Both the $n_{\rm H_2}$ and $T_{\rm kin}$ values for the cores with the SO non-detections, Per264, Per615, Per746 and Per799, are within the range of other cores where SO is detected. Moreover, these cores come from different regions within the Perseus molecular cloud; Per264 from NCG1333, Per615 and Per746 from IC348, and Per799 from B5, some of these with protostellar feedback. In this case, neither the physical properties of the cores nor their environments seem to be the sole reason responsible for the non-detections of SO.  \

\begin{figure}[h]
\centering
\includegraphics[width=8cm]{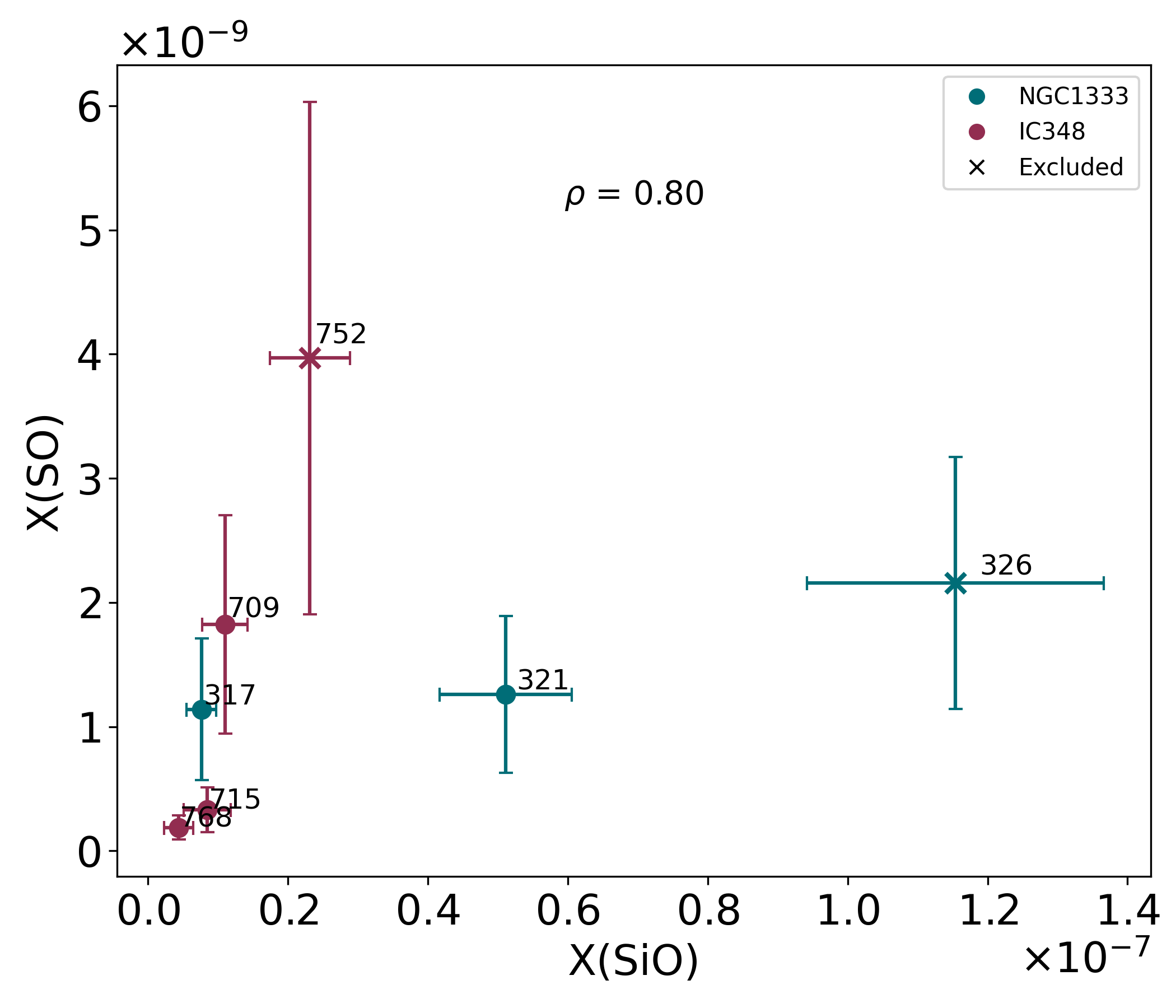}
\caption{Comparison of SO and SiO abundances. The values for SiO are taken from \cite{scibelli:25a}. The cores belonging to the NGC1333 region are plotted in teal, while the cores in IC348 are plotted in red. The $\rho$ including all the cores, except excluded cores plotted with a cross, is shown in the top centre of the figure.}
\label{sovssio}
\end{figure}

While some cores lack detections, others stand out due to their notably high column densities and abundances. Per326 shows particularly high HCS$^{+}$, OCS, H$_{2}$CS and SO column densities, of an order of a magnitude higher compared to the other cores in the sample. Per326 also has the highest column densities for CH$_{3}$OH, CH$_{3}$CHO, H$_{2}$CCO, HCOOCH$_{3}$, CS, $^{13}$CS, C$^{34}$S and HC$^{13}$CCCCN \citep{scibelli:24, pokorny:26}. Moreover, Per326 is the core with the brightest SiO emission, and the only core in the sample where PN, PO and PO$^{+}$ are detected \citep{scibelli:25a}. The detection of the phosphorous-bearing molecules alongside SiO is suggested to be a sign of shock chemistry, possibly driven by the outflow of the nearby SVS13 A protostar. SiO has been detected towards Per264, Per317, Per321 and Per326 of the NGC1333 region and cores Per709, Per715 and Per752 from the IC348 region. These cores are in the vicinity of protostellar feedback, which could link the presence of SiO to mild shocks from outflows. If we plot SO vs SiO we get a positive strong correlation (Figure \ref{sovssio}, $\rho$ = 0.80). The cores in the NGC1333 and IC348 regions appear to have strong correlations ($\rho$ = 1.00) with different slopes with respect to each other. Nevertheless, due to the finite number of points, specially for NGC1333, the correlations within the same region can not be confirmed. SO is also detected towards cores in our sample which are not in the vicinity of protostellar feedback, and is not detected towards Per264 which is. Therefore, the SO abundance may be enhanced by mild shocks in cores located in the vicinity of protostellar outflows, but is not entirely dependent on these interactions. Even if Per326 has the highest HCS$^{+}$, OCS, H$_{2}$CS and SO column densities, their abundances are within the spread of observed values towards all cores. This is due to Per326 being the densest source from the sample.\

Past works point at Per752 having a peculiar chemical composition with the presence of COMs, such as CH$_{3}$OH and CH$_{3}$CHO \citep{scibelli:24}, but non-detections of relatively large carbon chains such as c-C$_{3}$H$_{2}$, CCCS, HC$_{5}$N and HC$_{7}$N \citep{ferrerasensio:26, pokorny:26}. This suggests that Per752 possesses a rich overall chemistry that is selectively deficient in carbon chains.\

In the present study, Per752 presents some particularities related to its sulphur chemistry compared to the rest of the sample. Source Per752 stands out for having a remarkably high OCS column density, the highest after Per326, while having an average SO column density, and the lowest HCS$^{+}$ and H$_{2}$CS column densities of the sample. Per752 has the highest OCS and SO abundances, with values 1.3$\times$10$^{-9}$ and 4.0$\times$10$^{-9}$, respectively. On the other hand, the HCS$^{+}$ and H$_{2}$CS abundances are average. It needs to be kept in mind that the SO abundance has a large uncertainty, (4.0 $\pm$ 2.1) $\times$ 10$^{-9}$, and agrees, within these limits, with other high-abundance-SO cores such as Per321, Per326 and Per709. Nevertheless, Per752 remains one of the cores with the highest SO abundance.\ 

Due to the $X_{\rm OCS}$ and $X_{\rm SO}$ outlier nature Per326 and Per752 are not included in the correlation analyses (see Section \ref{physs}). Despite its low volume density, 2.2$\times$10$^{4}$ cm$^{-3}$, usually linked to early age, the combination of COM detections, high SO and OCS, and low carbon chain abundances, point to an advanced degree of chemical maturity in Per752. OCS and SO are considered to be late-type molecules \citep{ruffle:99, wakelam:04}. An increase of the OCS abundance with a peak at 10$^{6}$ years is seen for the TMC-1 astrochemical model in \cite{laas:19}. Moreover, \cite{ghosh:24} reports an increase of the OCS abundance from the prestellar to the protostellar Class 0 phase by observations towards L1544 (6.4$\times$10$^{-11}$), B1-b (2.5$\times$10$^{-10}$) and IRAS4A (1.4$\times$10$^{-9}$). In this work we find OCS abundances with an average of 3.2$\times$10$^{-10}$ for all cores, with Per752 having an abundance of $\sim$10$^{-9}$. The SO abundance is seen to peak later, 3.6$\times$10$^{5}$ yrs, compared to carbon-bearing sulphur species like CS, $\sim$2.5$\times$10$^{5}$ yrs, in the astrochemical model presented in \cite{vidal:17}. In fact the SO/CS ratio is often used as a chemical clock (\citealt{ruffle:99, esplugues:23}, see Section \ref{env}).  Lastly, in a similar way than OCS, the abundance of SO also has a tendency to increase from the prestellar core stage to the Class 0 protostellar stage \citep{ghosh:24}. It needs to be kept in mind that chemical maturity is not necessarily a consequence of an advanced evolutionary state of the core, but can be a consequence of other physical, chemical or environmental factors.\

\subsection{Environmental and core property influences}\label{env}

The relative abundances of the molecules studied in this work vary across cores. This indicates that the local characteristics and environments of the different cores result in different molecular production rates and even desorption rates. The spread in column density for the same molecule across the core sample can be sorted from the smallest to the largest spread as H$_{2}$CS, HCS$^{+}$, OCS and SO. The abundance spreads for H$_{2}$CS, HCS$^{+}$, OCS and SO, which are calculated by subtracting the largest and smallest abundances measured across the core sample, are 5.4$\times$10$^{-10}$, 6.0$\times$10$^{-11}$, 6.0$\times$10$^{-10}$ and 37.8$\times$10$^{-10}$, respectively. This result hints at an increasing dependence on the local environment for these molecules respectively.\ 

Some cores in this work, Per264, Per326, Per709 and Per715, are on the path of protostellar outflows on the NGC 1333 and IC 348 regions \citep{gueth:99, plunkett:13}. Moreover, Per326 is in a post-shock gas region, traced by SiO \citep{lefloch:98}, and as mentioned in the previous section, is the only core in the sample that shows emission of PN, PO and PO$^{+}$ which are linked to shock chemistry \citep{scibelli:25a}. Shocks are known to enhance the abundance of sulphur-bearing molecules \citep{martin:92, pineau:93, chernin:94, bachiller:97, esplugues:14, bouscasse:22}. Nevertheless, in this study higher abundances of sulphur-bearing molecules are not consistently found towards these cores. Consequently, based on the limited number of transitions observed, we find no definitive evidence of sulphur chemistry being enhanced by nearby shock activity. However, we cannot entirely rule out localized shock enhancement. If the shocked gas is confined to small, unresolved regions, the resulting emission could be significantly beam-diluted relative to the more extended ambient core emission.\

In order to see the influence of the physical properties of the cores in the abundance of the HCS$^{+}$, OCS, H$_{2}$CS and SO molecules, these molecular abundances are plotted against $n_{\rm H_2}$ and $T_{\rm kin}$ (Figure \ref{phys}). It needs to be kept in mind that molecular abundances depend on a lot of factors, and consequently, the following discussions are done in a broad and qualitative way.\

A very strong inverse correlation is found for $X_{\rm HCS^{+}}$ with $n_{\rm H_2}$ ($\rho$ = -0.82), a strong direct correlation is found between $X_{\rm SO}$ and $n_{\rm H_2}$ ($\rho$ = 0.73) and a strong inverse correlation is found for $X_{\rm H_{2}CS}$ with $n_{\rm H_2}$ ($\rho$ = -0.70). Nevertheless, the strength of the correlation between $X_{\rm H_{2}CS}$ and $n_{\rm H_2}$ depends on the inclusion of Per752, and if included, the $\rho$ is -0.53, which does not qualify as a strong correlation. These can be interpreted, if we consider $n_{\rm H_2}$ to be directly related to the evolutionary state of the core, as HCS$^{+}$ being more abundant in young cores, and SO being more abundant towards evolved cores. In fact, HCS$^{+}$ is considered an early-type molecule. It is formed from S$^{+}$ which is most abundant in translucent mediums, linked to young age or external layers of dense cores, through the reaction S$^{+}$ + CH $\rightarrow$ CS$^{+}$ + H$_{2}$ $\rightarrow$ HCS$^{+}$ \citep{vidal:17}. \cite{nagy:19} found higher abundances of CCS, HCS$^{+}$, C$^{34}$S and C$^{33}$S with respect to methanol by factors of $\sim$2-20, towards the L1521E starless core compared to the prestellar core L1544. The authors suggest that the more evolved source, L1544, shows lower sulphur-bearing molecule abundances as these may have already frozen out onto the surface of dust grains. Nevertheless, $n_{\rm H_2}$ represents an average density of the core and therefore may not reliably trace its evolutionary stage, as cores with similar mean densities can be at different stages of evolution \citep{moon:25a, moon:25b, moon:25c}. In contrast, the central density more directly probes evolutionary time \citep{tokuda:20, moon:25c, andras:26}. However, the central density can exhibit significant temporal variability due to dynamical evolution, including oscillatory or stochastic behaviour \citep{broderick:07, keto:14b, offner:25}, which may be partially smoothed out when considering the average density. \

On the other hand, SO is considered a late-type molecule, as already introduced above. Due to atomic carbon depletion happening before, at $\sim$10$^{5}$ yr, than atomic oxygen depletion, at $\sim$8$\times$10$^{5}$ yr, more oxygen is available for reactivity at later times thus enhancing the abundance of oxygen-bearing molecules in more evolved cores \citep{vidal:17}. The increase of SO abundance with evolutionary time within the starless/prestellar core stage is supported by extrapolating results from \cite{ghosh:24} where increasing SO abundance is observed from prestellar core to Class 0 protostars.\

Nevertheless, in \cite{ghosh:24} a similar trend is seen for OCS, and in this study a moderate inverse correlation for the latter molecule is found with $n_{\rm H_2}$. Moreover, in \cite{nagy:19} the OCS abundance is seen to decrease from the starless (L1521E) to the prestellar (L1544) phase, supporting the trend found in this work. The discrepancy between the OCS trend seen in this study and \cite{nagy:19} compared to the trend seen in \cite{ghosh:24} could be due to the molecular abundances not necessarily changing monotonically with time. In the starless and prestellar stages studied in \cite{nagy:19} and in this work, OCS increasingly depletes onto dust grains as the sources evolve toward higher densities and lower temperatures. In the subsequent stages of the star formation process, as the temperature increases due to the presence of an embedded protostar, the OCS trapped in the icy mantles of dust grains is released back into the gas phase, driving a significant increase in its gas-phase abundance relative to the cold prestellar phase, as observed by \cite{ghosh:24}. This increase of OCS abundance with time is also seen towards hot cores in \cite{viti:99}. Moreover, the difference in trends between SO and OCS, which are both considered late-type molecules (e.g., \citealt{santos:24}), in the starless stage seen in this work could be due to the first being formed primarily on the gas phase, and the latter being formed efficiently both in the gas and solid phases \citep{hasegawa:93, adriaens:10, laas:19, santos:24}. The moderate inverse correlations of the H$_{2}$CS could also be explained by this molecule being depleted over time.\ 

\cite{rodriguez:21} conducted similar HCS$^{+}$, OCS and SO abundance comparisons with $n_{\rm H_2}$ for an extended sample of starless cores towards the Perseus, Taurus and Orion molecular clouds within the Gas phase Elemental Abundances in Molecular cloudS (GEMS) IRAM 30m large program. In their study, they found an inverse correlation between the abundance of SO and $n_{\rm H_2}$ (see Figure 15 in \citealt{rodriguez:21}), contrarily to what is found in this paper. Moreover, no strong trends are found between HCS$^{+}$ and OCS abundances and $n_{\rm H_2}$. The lack of correlation between HCS$^{+}$ and $n_{\rm H_2}$ in \cite{rodriguez:21} also contrasts with the results shown in the present study. A possible explanation for the discrepancies between \cite{rodriguez:21} and the present study can be attributed to the different beams of the observations tracing different material affected by different physical conditions. The beam of the Yebes 40m observations in the present work ($\sim$36$^{\prime\prime}$--54$^{\prime\prime}$) may be picking up emission from the outer layers of the cores the beam of the GEMS data observed with the IRAM 30m telescope ($\sim$15$^{\prime\prime}$--29$^{\prime\prime}$) is not. Another possible cause is the core samples not being large enough to derive robust statistics. Lastly, it also needs to be kept in mind that the different SO and OCS transitions are observed in this work and in \cite{rodriguez:21} likely trace gas at different conditions.\ 

\begin{figure}[h]
\centering
\includegraphics[width=8cm]{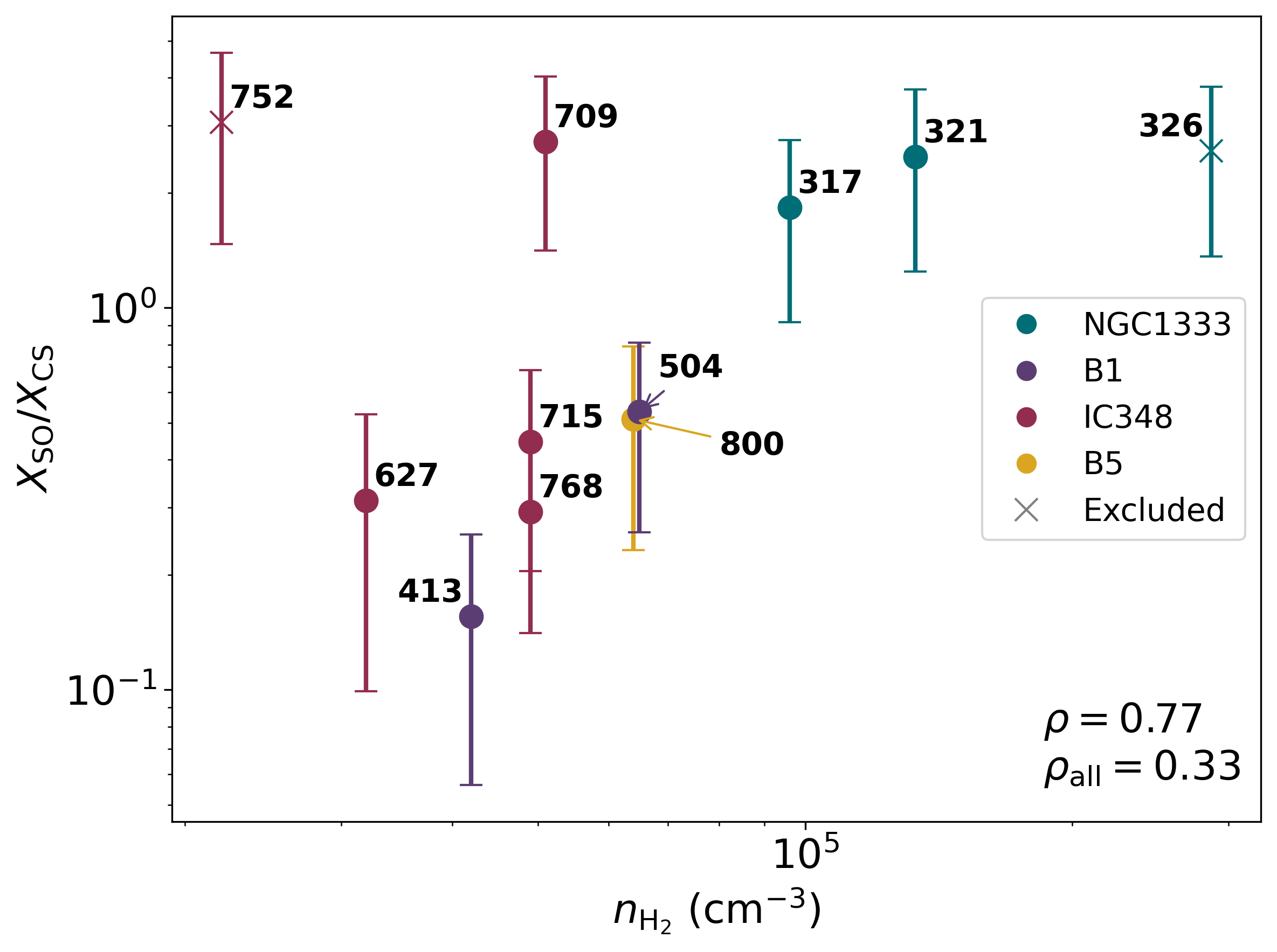}
\caption{The abundance ratio of SO and CS is plotted against the volume density of the cores. The CS abundances have been calculated using the observations in \cite{pokorny:26}. Both the $x$ and $y$ axes are plotted in a logarithmic scale for visualisation purposes. Both the Spearman rank coefficient excluding cores Per326 and Per752, plotted with a cross symbol, ($\rho$), and including all of the cores ($\rho_{all}$), are plotted on the bottom right corner.}
\label{socs}
\end{figure}

\begin{figure}[h]
\centering
\includegraphics[width=8cm]{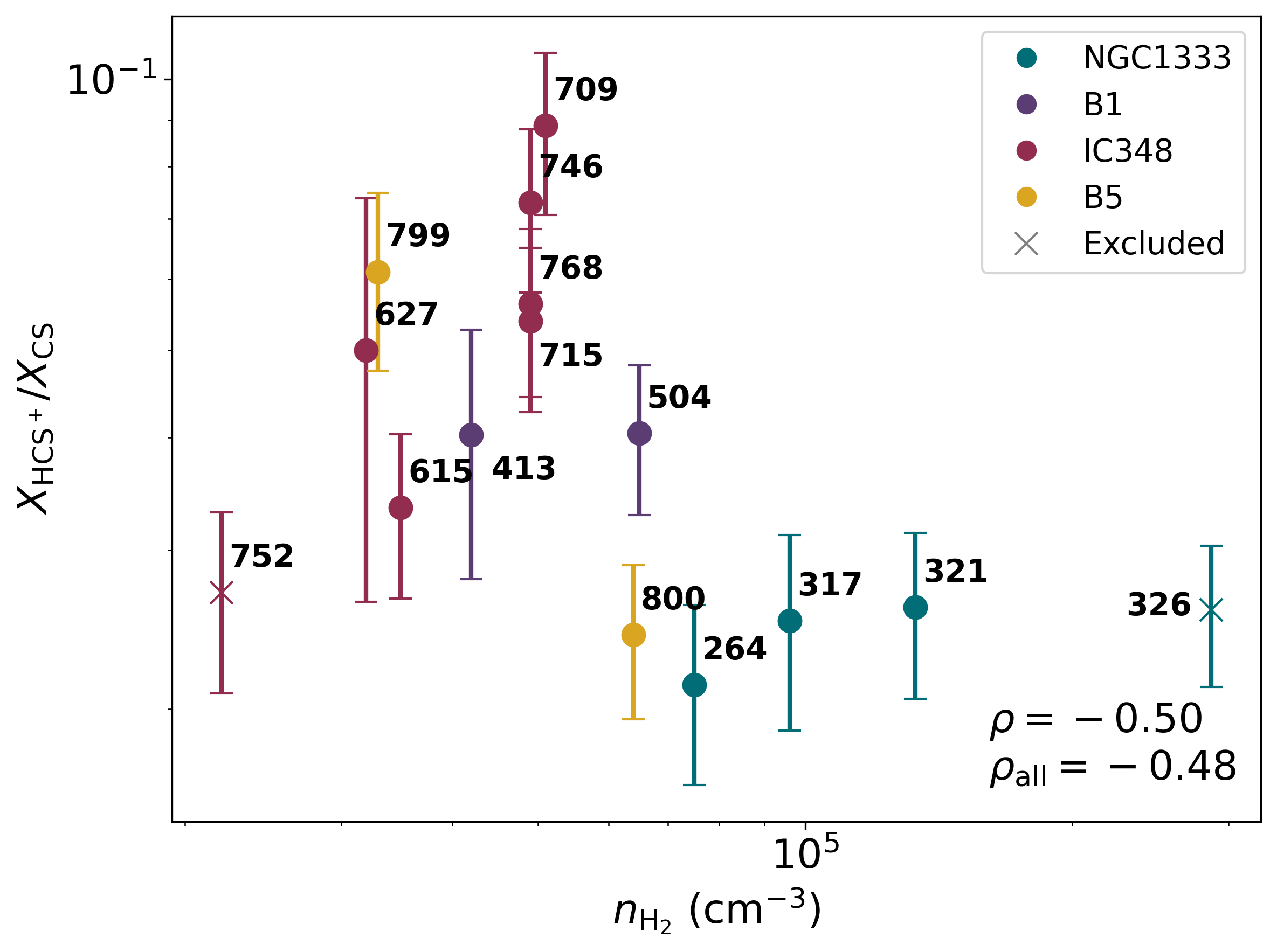}
\caption{Same as for Figure \ref{socs} but for the abundance ratio of HCS$^{+}$ to CS.}
\label{hcscs}
\end{figure}

Additionally, the SO/CS and HCS$^{+}$/CS abundance ratios are plotted against the $n_{\rm H_2}$ for each core in Figures \ref{socs} and \ref{hcscs}. A low SO/CS ratio is related to young cores, while a high ratio is related to older cores. Cores Per317, Per321, Per752, Per326 and Per709 have high uncertainties. Excluding core Per752, due to its SO abundance outlier behaviour, a strong correlation ($\rho$ = 0.77), is retrieved. When using the Pearson correlation coefficient, which is more sensitive to outliers, a $r$ = 0.65 value is obtained instead. Including both outliers, Per326 and Per752, results in $\rho_{\rm all}$ = 0.33. This would indicate that the physical and chemical evolutionary stages of the cores match. Per752 is an exception, presenting a high degree of chemical maturity, traced by a high $X_{\rm SO}$/$X_{\rm CS}$ ratio, but a low volume density, 2.2$\times$10$^{4}$ cm$^{-3}$. On the other hand a high HCS$^{+}$/CS ratio is related to younger cores, due to HCS$^{+}$ being an early type molecule. A mild inverse correlation is found between $X_{\rm HCS^+}$/$X_{\rm CS}$ and $n_{\rm H_2}$ ($\rho$ = -0.50). The Spearman rate coefficient obtained when including the outlier cores is -0.48. The SO/CS and HCS$^{+}$/CS abundance ratio direct and inverse tendencies, respectively, with $n_{\rm H_2}$ presented in this work agree with what has been found for other starless cores as well as Class 0 sources (see Figure 12 in \citealt{esplugues:23} and Figure 6 in Fern{\'a}ndez-Ru{\'i}z et al., accepted).    \

Strong inverse correlations are found for $X_{\rm OCS}$ and $X_{\rm H_{2}CS}$ with $T_{\rm kin}$, with $\rho$ = -0.71 and -0.75, respectively. The inverse correlation of $X_{\rm OCS}$ and $X_{\rm H_{2}CS}$ with $T_{\rm kin}$ points at OCS and H$_{2}$CS being more efficiently formed at lower temperatures or destroyed more effectively at higher temperatures. Both OCS and H$_{2}$CS can effectively be formed on the surface of dust grains upon depletion, which is favoured at lower temperatures \citep{laas:19, esplugues:22}. Nevertheless, the derived $T_{\rm kin}$ may be tracing the external layers of the prestellar cores rather than the temperatures at their centre. These sulphur-bearing species abundances are seen to peak in the external layers of cores in chemical models \citep{vastel:18}.   \

The $X_{\rm SO}$ has a moderate positive correlation with $T_{\rm kin}$, which could mean that an increase of the temperature could mildly promote the formation of SO. Lastly, $X_{\rm HCS^{+}}$ has a weak correlation with $T_{\rm kin}$. Generally, the abundances of HCS$^{+}$ and SO seem to depend more on the evolutionary state of the core, traced by volume density, while OCS and H$_{2}$CS seem to be more affected by the kinetic temperature. \

\cite{rodriguez:21} does not find a strong correlation between the OCS abundance and the $T_{\rm kin}$ in a sample of starless cores in the Perseus and Taurus molecular clouds. The possible explanations for the discrepancy are, as introduced above for the comparisons with $n_{\rm H_2}$, due to different beam sizes, limited core sample or a bias introduced by the molecular transitions analysed.\

In addition to density and temperature, the chemical evolution of starless cores is heavily influenced by external environmental factors, particularly the far-ultraviolet (FUV) component of the ambient interstellar radiation field (ISRF). While the lower-energy visible and infrared ISRF regulates core temperatures and grain-surface freeze-out \citep{evans:01, caselli:12}, the FUV photons directly govern the ionization fractions and photodissociation rates of molecular tracers along the core boundaries \citep{hollenbach:99}. However, low-energy cosmic rays (LECRs) are unique in their ability to penetrate deep into the densest regions of molecular cores. Recent high-resolution HI narrow self-absorption (HINSA) mapping in the Perseus molecular cloud has revealed a significant environmental disparity in these ionization fields that was previously hidden at macroscopic scales \citep{luo:26}. Specifically, star-forming regions like IC348 exhibit cosmic ray ionization rate (CRIR) values nearly an order of magnitude higher than those in the neighbouring NGC 1333 (see also \citealt{pineda:24b}), a contrast likely driven by local acceleration sources such as stellar-wind termination shocks. Particularly, the location of Per752 shows the highest HINSA fraction towards both IC348 as well as NGC1333, which could partially explain the unique chemical print of this core (Figure 1, panel a in \citealt{luo:26}). A high CRIR could enhance the desorption of frozen out molecules in the surface of grains such as SO or OCS, while destroying gas-phase carbon chains.\

Finally, another important factor to consider when studying chemical abundances is turbulence, since it may induce density variations which play an important role in shaping the chemistry of molecular clouds. This is especially relevant for those molecules sensitive to hydrogen density, such as some sulphurated species \cite{beitia:24}.

\subsection{Chemical links between sulphur-bearing molecules}\label{chem}

With the aim of exploring the chemical relationship between the different molecules, the abundances of pairs of molecules are plotted against each other in Figure \ref{triangle}. The molecule whose abundance is found to have the highest correlation with other molecular abundances is CS. Due to the extremely low reaction coefficient $k$(T), of the H$_{2}$ + S$^{+}$ $\rightarrow$ H$_{2}$S$^{+}$ + $\nu$, $k$(10 K) = 1.97$\times$10$^{-17}$ cm$^{-3}$ \citep{vidal:17}, and the higher ionization potential of hydrogen, the sulphur ion reacts more readily with carbon than hydrogen, producing CS. Consequently, CS is one of the most important precursors of sulphur chemistry \citep{vastel:18}. The molecular abundance correlation of CS is the strongest with H$_{2}$CS ($\rho$ = 0.92). These molecules are directly related by the destruction of H$_{2}$CS which produces CS, H$_{2}$CS + C $\rightarrow$ CH$_{2}$ + CS. Moreover, both of them form from the reaction of S$^{+}$ with a small hydrocarbon precursor, CH for CS and CH$_{4}$ for H$_{2}$CS (for further details see \citealt{laas:19}). Reactions with S$^{+}$ are expected to be more relevant in lower-density regions, such as younger cores or the outer layers of dense cores.\

The CS abundance is also directly correlated with CCS and CCCS ($\rho$ = 0.91 and 0.89). CCS is thought to be formed in the surface of grains directly from CS through the neutral-neutral reaction CH + CS $\rightarrow$ H + CCS. On the gas phase CCS is formed from the S$^{+}$ + C$_{2}$H combination, followed by a chain of hydrogenation and electronic recombination reactions, similarly to CS \citep{sakai:07, laas:19}. The formation of CCCS follows similar reactions than for CCS, including additional reactions where it is formed directly from CS, C$_{2}$H + CS $\rightarrow$ CCCS + H \citep{laas:19}.\

The abundance relationship between CS and OCS is $\rho$ = 0.78. OCS can be formed both in the gas phase as well as on the surface of dust grains. On the gas phase one of the reactions that contribute to its formation is the CS + OH $\rightarrow$ H + OCS, which directly relates CS and OCS \citep{laas:19}. The destruction of OCS through carbon also yields CS. Moreover, the reaction of CS and O atoms in the surfaces of grains can also yield OCS \citep{vastel:18}.\

A moderate correlation is found between CS and HCS$^{+}$. These molecules are directly related by the electronic recombination of HCS$^{+}$ into CS, although this specific reaction pathway only partially accounts for the total CS production \citep{vidal:17}. CS can also react with H$_{3}^{+}$, HCO$^{+}$ and H$_{3}$O$^{+}$ to give HCS$^{+}$.\

Generally, molecules that have a high degree of correlation with CS also show correlation amongst each other. This can be interpreted as CS being a central piece in the sulphur chemical network. Finally, CS weakly inversely correlates with SO ($\rho$ = -0.45). This could be due to their different formation timescales, being CS formed in the early stages and SO forming later on and the fact that they are involved in different chemical networks, the carbon and the oxygen one, respectively.\

As a matter of a fact, SO shows inverse correlations with the rest of the molecules, hinting at this molecule following a different chemistry. As suggested before in the text, this could be interpreted as SO being a late-type molecule in contrast with the others. Nevertheless, OCS is also commonly considered a late-type molecule and its $\rho$ value with SO is -0.52, which indicates opposite trends, when a direct relation should be expected. Nevertheless, this anticorrelation is moderate, so simple conclusions can not be made. As discussed in Section \ref{env}, molecular abundances may not change monotonically with time resulting in some of the behaviours seen.\

\subsection{Comparison with the Taurus molecular cloud}\label{taurus}

A recent sulphur-bearing molecule survey towards nine starless and prestellar cores in the Taurus molecular cloud includes observations of CS, CCS, CCCS, OCS, SO and H$_{2}$CS \citep{schoeller:26}. Similarly to this work, \cite{schoeller:26} finds a range of molecular abundances amongst the sources. The abundance for SO is computed from the 2$_{3}$--1$_{2}$ transition for all sources except for source L1495B where they use the 2$_{2}$--1$_{1}$, the abundance for OCS is computed from the 8--7 transition, the one for H$_{2}$CS from the 3$_{1,2}$--2$_{1,1}$, and the one for HCS$^{+}$ from the 2--1 transition.  The SO abundances range between 7.2 $\pm$ 1.3$\times$10$^{-11}$ and 7.7 $\pm$ 1.4$\times$10$^{-10}$, for OCS 1.1 $\pm$ 0.2$\times$10$^{-10}$ and 5.4 $\pm$ 1.1$\times$10$^{-10}$, for H$_{2}$CS 3.7 $\pm$ 0.7$\times$10$^{-11}$ to 6.4 $\pm$ 1.3$\times$10$^{-10}$, and for HCS$^{+}$ 0.9 $\pm$ 1.0$\times$10$^{-11}$ to 6.4 $\pm$ 4.2$\times$10$^{-11}$. The average abundances are for SO, 4.0$\times$10$^{-10}$, for H$_{2}$CS 2.5$\times$10$^{-10}$, for OCS 3.0$\times$10$^{-10}$ and for HCS$^{+}$ 3.6$\times$10$^{-11}$. The average abundances for HCS$^{+}$, OCS, and H$_{2}$CS towards Taurus in \cite{schoeller:26} and Perseus in the present study agree within a factor of $\sim$1.1. Nevertheless, the average SO abundance in Perseus is $\sim$2.7 higher than the one found in Taurus. This could be explained by the average SO abundance in Perseus being affected by exceptionally high SO abundances of some cores, such as Per752.   \

No strong correlations are found for the molecules and core evolutionary tracers such as the H$_{2}$ column density, the N$_{2}$D$^{+}$/N$_{2}$H$^{+}$ ratio and the CO depletion factor in \cite{schoeller:26}. They attribute these results to the fact that sulphur-bearing molecules trace the outer regions of cores, while the evolutionary tracers tested trace the central regions of cores. Nevertheless, in this study, we find a strong direct correlation between the abundance of SO with $n_{\rm H_2}$ volume density ($\rho$ = 0.73), a strong inverse correlation between the HCS$^{+}$ and H$_{2}$CS abundances and $n_{\rm H_2}$ ($\rho$ = -0.82, -0.70, respectively), even if the H$_{2}$CS correlation is deemed less robust (see Section \ref{physs}), and a moderate inverse correlation between the abundance of OCS with $n_{\rm H_2}$ ($\rho$ = -0.44). The differences in trends could be due to the fact that volume density is a more direct physical driver of chemical processes like freeze-out than line-of-sight integrated column density.\

\cite{schoeller:26} further examines chemical trends by plotting the C$^{34}$S/$^{34}$SO and the CCS/$^{34}$SO abundances against the N$_{2}$D$^{+}$/N$_{2}$H$^{+}$ ratio ($\rho$ = -0.86 and -0.81, respectively). These results agree with what is found in this study. In the present work this trend is tested by plotting the fraction of SO and CS abundances with $n_{\rm H_2}$, retrieving a strong direct correlation, $\rho$ = 0.77. Notice that in \cite{schoeller:26} the C/O fraction is given, while in this work the O/C fraction is given. The agreement between $N_{\rm H_2}$ and $n_{\rm H2}$ for these ratios, despite the disagreement for individual molecules, suggests that while absolute abundances are sensitive to local environmental variations and spatial structure, the relative shift from carbon-rich to oxygen-rich chemistry is a more robust indicator of global chemical maturity. Ultimately, these two works agree that sulphur-bearing molecule abundances are dependent on multiple factors, including physical structure and local environment.\

\section{Conclusions}\label{conclusions}

New HCS$^{+}$, OCS, H$_{2}$CS and SO observations towards fifteen starless and prestellar cores in the Perseus molecular cloud taken with the Yebes 40m telescope are presented in this work. The molecules are detected in the 100\%, 93\%, 100\% and 73\% of the sample, respectively. The non-detections do not seem to be directly related to the physical properties of the cores nor the regions within the Perseus molecular cloud they belong to. The column densities for all the core sample computed assuming LTE, are for HCS$^{+}$ (2.2 - 17.0)$\times$10$^{12}$ cm$^{-2}$, for OCS (1.9 - 25.0)$\times$10$^{12}$ cm$^{-2}$, for H$_{2}$CS (1.0 - 15.0)$\times$10$^{12}$ cm$^{-2}$ and for SO (0.3 - 17.0)$\times$10$^{13}$ cm$^{-2}$.\

The abundances of the studied sulphur-bearing molecules are not consistently higher towards cores in the path of protostellar outflows, as could have been expected from past literature evidence of the enhancement of sulphur chemistry in these environments \citep{pineau:93}. The abundance of OCS and SO are particularly high, with respect to the rest of the sample, for core Per752. The high abundance of the OCS and SO late-type molecules, combined with the presence of COMs and the low abundance of carbon chains in this core reported in the literature, points to a particularly evolved chemical signature. Nevertheless, its low volume density does not support a physically mature state, indicating that this advanced chemical stage might be a consequence of factors other than chronological age such as the enhanced cosmic ray ionization rate observed towards this core position.

The SO/CS abundance ratio, considered a chemical clock, correlates well ($\rho$ = 0.77) with the core volume density, indicating that the physical and chemical evolutionary times of these sources match. The exception is Per752, which shows a high SO abundance compared to its low volume density.\

When plotting the molecular abundances against the $n_{\rm H_2}$ and $T_{\rm kin}$ core values inverse correlations are found for HCS$^{+}$, OCS and H$_{2}$CS and direct correlations for SO. HCS$^{+}$, H$_{2}$CS and SO have strong inverse and direct correlations respectively, while OCS shows a mild inverse correlation with $n_{\rm H_2}$. These behaviours have been linked to the time where these abundances peak combined with their depletion on the surface of grains. On the other hand, OCS and H$_{2}$CS show strong inverse correlations, while HCS$^{+}$ shows a weak inverse correlation and SO shows a mild direct correlation with $T_{\rm kin}$. The abundance of HCS$^{+}$ and SO are more dependent on $n_{\rm H_2}$ and OCS and H$_{2}$CS are more dependent on $T_{\rm kin}$.\

All the molecules show direct correlations between each other with the exception of SO, which presents inverse correlations with the rest. This can be explained by the late-type character of SO combined with its differentiated ``oxygen chemistry'' versus the ``carbon chemistry'' of the rest, in a similar fashion to that of nitrogen chemistry (see \citealt{megias:23}). In these results OCS align more with the carbon-bearing molecules than with SO. The molecule that shows the strongest correlations with others is CS, as expected from its central role in sulphur chemistry.\

\begin{acknowledgements}
J. Ferrer Asensio thanks RIKEN Special Postdoctoral Researcher Program (Fellowships) for financial support. S. Scibelli acknowledges the National Radio Astronomy Observatory which is a facility of the National Science Foundation operated under cooperative agreement by Associated Universities, Inc. L. Steffes is supported by a National Science Foundation Graduate Research Fellowship Program under Grant \textnumero [DGE-2137419]. Any opinions, findings, and conclusions or recommendations expressed in this material are those of the authors and do not necessarily reflect the views of the National Science Foundation. L. Steffes and Y. Shirley are also supported by the National Science Foundation Astronomy and Astrophysics Grant (AAG) AST-2205474. I. Jim{\'e}nez-Serra and A. Meg{\'i}as acknowledge funding from the ERC grant OPENS (GA No. 101125858) funded by the European Union, and from the grant PID2022-136814NB-I00 funded by the Spanish Ministry of Science, Innovation and Universities / State Agency of Research, MCIU/AEI/10.13039/501100011033 and by ``ERDF/EU''. This work is supported by ERC grant SUL4LIFE, GA No. 101096293 funded by the European Union. A.F. and G. E. thank project PID2022-137980NB-I00 funded by the Spanish Ministry of Science and Innovation/State Agency of Research MCIN/AEI/10.13039/501100011033 and by “ERDF A way of making Europe”.
\end{acknowledgements}

\bibliographystyle{aa}
\bibliography{AA5860425}

\appendix

\onecolumn

\section{Observed transitions}\label{obstran}

The detected HCS$^{+}$, OCS, H$_{2}$CS and SO transitions towards the Perseus molecular cloud sample are presented in Table \ref{obstrans}. The peak temperature ($T_{\rm MB}$), position velocity ($v_{\rm LSR}$), width (FWHM), rms and signal-to-noise ratio (S/N) as calculated with the peak intensity, of the lines are given. The SO 3$_{2}$--2$_{2}$ transition for core Per627 was not well fitted with ``pyspeckit''. For this line, marked with an asterisk ($*$), the integrated intensity and the assumed FWHM are written instead.\

\begin{longtable}{lccccccc}
\caption{Detected HCS$^{+}$, OCS, H$_{2}$CS and SO transitions.}\\
\hline\hline
Core number & Molecule & Transition & $T_{\rm MB}$ & $v_{\rm LSR}$ & FWHM & rms & S/N \\
 &  &  & (mK) & (km/s) & (km/s) & (mK) &  \\
\hline\hline
264 & HCS$^{+}$ & 1--0 & 28 & 8.028 (0.009) & 0.596 (0.021) & 3 & 9 \\
 & OCS & 3--2 & 22 & 7.937 (0.015) & 0.691 (0.036) & 4 & 6 \\
 & OCS & 4--3 & 29 & 7.887 (0.011) & 0.802 (0.027) & 8 & 3 \\
 & H$_{2}$CS & 1$_{0,1}$--0$_{0,0}$ & 42 & 8.053 (0.009) & 0.929 (0.021) & 2 & 22 \\
\hline
317 & HCS$^{+}$ & 1--0 & 30 & 8.312 (0.010) & 0.743 (0.023) & 4 & 8 \\
 & OCS & 3--2 & 18 & 8.349 (0.015) & 0.652 (0.038) & 2 & 10 \\
 & OCS & 4--3 & 42 & 8.456 (0.007) & 0.743 (0.016) & 9 & 4 \\
 & H$_{2}$CS & 1$_{0,1}$--0$_{0,0}$ & 35 & 8.431 (0.010) & 0.865 (0.023) & 2 & 16 \\
 & SO & 3$_{2}$--2$_{2}$ & 79 & 8.255 (0.004) & 0.696 (0.009) & 4 & 19 \\
\hline
321 & HCS$^{+}$ & 1--0 & 26 & 8.542 (0.014) & 1.011 (0.033) & 2 & 10 \\
 & OCS & 3--2 & 22 & 8.533 (0.017) & 0.975 (0.040) & 2 & 12 \\
 & OCS & 4--3 & 45 & 8.598 (0.007) & 0.869 (0.016) & 6 & 7 \\
 & H$_{2}$CS & 1$_{0,1}$--0$_{0,0}$ & 48 & 8.558 (0.008) & 0.889 (0.018) & 2 & 30 \\
 & SO & 3$_{2}$--2$_{2}$ & 95 & 8.438 (0.004) & 0.879 (0.009) & 3 & 28 \\
\hline
326 & HCS$^{+}$ & 1--0 & 87 & 7.922 (0.004) & 1.064 (0.010) & 3 & 31 \\
 & OCS & 3--2 & 101 & 7.655 (0.004) & 1.059 (0.009) & 2 & 45 \\
 & OCS & 4--3 & 180 & 7.636 (0.002) & 0.943 (0.004) & 7 & 25 \\
 & H$_{2}$CS & 1$_{0,1}$--0$_{0,0}$ & 131 & 7.900 (0.003) & 1.174 (0.008) & 1 & 87 \\
 & SO & 3$_{2}$--2$_{2}$ & 261 & 7.545 (0.001) & 1.144 (0.004) & 2 & 108 \\
\hline
413 & HCS$^{+}$ & 1--0 & 64 & 7.734 (0.005) & 0.733 (0.012) & 2 & 26 \\
 & OCS & 3--2 & 19 & 7.686 (0.017) & 0.765 (0.040) & 3 & 6 \\
 & OCS & 4--3 & 33 & 7.760 (0.008) & 0.634 (0.018) & 7 & 4 \\
 & H$_{2}$CS & 1$_{0,1}$--0$_{0,0}$ & 118 & 7.862 (0.003) & 0.646 (0.006) & 2 & 75 \\
 & SO & 3$_{2}$--2$_{2}$ & 9 & 7.509 (0.032) & 0.664 (0.074) & 3 & 3 \\
\hline
504 & HCS$^{+}$ & 1--0 & 48 & 6.730 (0.007) & 0.770 (0.016) & 3 & 16 \\
 & OCS & 3--2 & 32 & 6.631 (0.011) & 0.777 (0.026) & 2 & 18 \\
 & OCS & 4--3 & 63 & 6.739 (0.004) & 0.677 (0.010) & 3 & 21 \\
 & H$_{2}$CS & 1$_{0,1}$--0$_{0,0}$ & 85 & 6.804 (0.004) & 0.786 (0.010) & 3 & 31 \\
 & SO & 3$_{2}$--2$_{2}$ & 19 & 6.519 (0.020) & 0.879 (0.047) & 2 & 10 \\
\hline
615 & HCS$^{+}$ & 1--0 & 50 & 8.147 (0.005) & 0.567 (0.012) & 3 & 17 \\
 & OCS & 3--2 & 32 & 8.111 (0.009) & 0.598 (0.020) & 2 & 13 \\
 & OCS & 4--3 & 47 & 8.153 (0.004) & 0.375 (0.011) & 13 & 4 \\
 & H$_{2}$CS & 1$_{0,1}$--0$_{0,0}$ & 81 & 8.223 (0.004) & 0.735 (0.010) & 3 & 31 \\
\hline
627 & HCS$^{+}$ & 1--0 & 48 & 8.476 (0.004) & 0.528 (0.011) & 4 & 11 \\
 & OCS & 3--2 & 17 & 8.376 (0.019) & 0.659 (0.043) & 2 & 7 \\
 & OCS & 4--3 & 25 & 8.437 (0.006) & 0.322 (0.029) & 8 & 3 \\
 & H$_{2}$CS & 1$_{0,1}$--0$_{0,0}$ & 54 & 8.506 (0.006) & 0.616 (0.013) & 2 & 32 \\
 & SO$^{*}$ & 3$_{2}$--2$_{2}$ & 12 & - & 0.500 & 3 & 6 \\
\hline
709 & HCS$^{+}$ & 1--0 & 46 & 8.913 (0.008) & 1.023 (0.019) & 4 & 12 \\
 & OCS & 3--2 & 17 & 8.745 (0.017) & 0.610 (0.039) & 2 & 11 \\
 & OCS & 4--3 & 32 & 8.810 (0.007) & 0.344 (0.014) & 6 & 5 \\
 & H$_{2}$CS & 1$_{0,1}$--0$_{0,0}$ & 22 & 8.876 (0.016) & 0.943 (0.039) & 2 & 13 \\
 & SO & 3$_{2}$--2$_{2}$ & 75 & 8.622 (0.004) & 0.647 (0.009) & 3 & 25 \\
\hline
715 & HCS$^{+}$ & 1--0 & 47 & 8.965 (0.006) & 0.666 (0.014) & 5 & 9 \\
 & H$_{2}$CS & 1$_{0,1}$--0$_{0,0}$ & 40 & 9.015 (0.009) & 0.736 (0.021) & 2 & 25 \\
 & SO & 3$_{2}$--2$_{2}$ & 10 & 8.767 (0.033) & 0.861 (0.077) & 2 & 5 \\
\hline
746 & HCS$^{+}$ & 1--0 & 45 & 8.784 (0.008) & 0.862 (0.018) & 4 & 13 \\
 & OCS & 3--2 & 14 & 8.902 (0.027) & 0.958 (0.063) & 3 & 5 \\
 & H$_{2}$CS & 1$_{0,1}$--0$_{0,0}$ & 47 & 8.896 (0.007) & 0.883 (0.018) & 2 & 26 \\

\newpage
\caption[]{Detected HCS$^{+}$, OCS, H$_{2}$CS and SO transitions (continued).} \\
\hline\hline
Core & Molecule & Transition & $T_{\rm MB}$ & $v_{\rm LSR}$ & FWHM & RMS & S/N \\
 & & & (mK) & (km/s) & (km/s) & (mK) & \\
\hline\hline
752 & HCS$^{+}$ & 1--0 & 19 & 7.838 (0.015) & 0.614 (0.036) & 3 & 6 \\
 & OCS & 3--2 & 43 & 7.507 (0.008) & 0.770 (0.018) & 2 & 19 \\
 & OCS & 4--3 & 82 & 7.498 (0.003) & 0.766 (0.008) & 8 & 11 \\
 & H$_{2}$CS & 1$_{0,1}$--0$_{0,0}$ & 13 & 7.801 (0.026) & 0.789 (0.060) & 2 & 6 \\
 & SO & 3$_{2}$--2$_{2}$ & 60 & 7.485 (0.006) & 0.762 (0.014) & 3 & 22 \\
\hline
768 & HCS$^{+}$ & 1--0 & 32 & 8.950 (0.011) & 0.876 (0.025) & 2 & 13 \\
 & OCS & 3--2 & 13 & 9.036 (0.024) & 0.672 (0.056) & 2 & 6 \\
 & H$_{2}$CS & 1$_{0,1}$--0$_{0,0}$ & 24 & 9.152 (0.014) & 0.827 (0.034) & 1 & 18 \\
 & SO & 3$_{2}$--2$_{2}$ & 9 & 9.010 (0.033) & 0.537 (0.063) & 1 & 9 \\
\hline
799 & HCS$^{+}$ & 1--0 & 37 & 10.137 (0.010) & 0.909 (0.024) & 3 & 11 \\
 & OCS & 3--2 & 40 & 10.093 (0.006) & 0.481 (0.018) & 3 & 13 \\
 & OCS & 4--3 & 68 & 10.144 (0.003) & 0.366 (0.006) & 7 & 10 \\
 & H$_{2}$CS & 1$_{0,1}$--0$_{0,0}$ & 72 & 10.250 (0.004) & 0.642 (0.010) & 2 & 44 \\
\hline
800 & HCS$^{+}$ & 1--0 & 57 & 10.264 (0.004) & 0.484 (0.010) & 3 & 18 \\
 & OCS & 3--2 & 50 & 10.183 (0.006) & 0.608 (0.013) & 3 & 19 \\
 & OCS & 4--3 & 87 & 10.248 (0.003) & 0.664 (0.007) & 7 & 12 \\
 & H$_{2}$CS & 1$_{0,1}$--0$_{0,0}$ & 90 & 10.299 (0.004) & 0.710 (0.009) & 4 & 23 \\
 & SO & 3$_{2}$--2$_{2}$ & 36 & 10.189 (0.008) & 0.561 (0.017) & 2 & 17 \\
\label{obstrans}
\end{longtable}

\twocolumn

\section{Column density calculation with RADEX}\label{columndensity}

The column densities are also calculated using the non-LTE approach with the RADEX code \citep{vandertak:07b}. As a consequence of limited constraints for the RADEX calculations, and due to the general agreement between the LTE and non-LTE approaches, the LTE-derived column densities are reported in the main text while the non-LTE results are presented separately in this section.\ 

RADEX is a non-LTE radiative transfer program that models molecular excitation. To do this, it requires physical parameters of the source, specifically the volume density ($n_{\rm H_2}$) and kinetic temperature ($T_{\rm kin}$), along with the molecule’s spectroscopic data and collisional rate coefficients. All of the molecules discussed in this paper have available collisional rate coefficients. These are extracted from both the Excitation of Molecules and Atoms for Astrophysics (EMAA)\footnote{\url{https://emaa.osug.fr}} database and the Leiden Atomic and Molecular Database (LAMDA)\footnote{\url{https://home.strw.leidenuniv.nl/~moldata/}}. The collisional rate coefficient files are based on \cite{schoier:05} (HCS$^{+}$), \cite{denisalpizar:23} (OCS), \cite{wiesenfeld:13} (H$_{2}$CS), and \cite{lique:06} (SO). The H$_{2}$CS molecule has separate ortho and para states. The transition observed in this study, 1$_{0,1}$--0$_{0,0}$, corresponds to the para state, and thus, the collisional rate coefficients for para-H$_{2}$CS are used. To compute the total H$_{2}$CS column densities a statistical ortho-to-para ratio of 3 is assumed. Column densities were estimated by modelling the line intensities using the hydrogen volume densities ($n_{\rm H_2}$) and kinetic temperatures ($T_{\rm kin}$) taken from \cite{scibelli:24} and listed in Table \ref{source}. These input values represent the median conditions derived from \textit{Herschel} maps \citep{pezzuto:21}, averaged over the 62$^{\prime\prime}$ beam of the Arizona Radio Observatory (ARO) 12m telescope, for the study in \cite{scibelli:24}. Kinetic temperatures for individual cores are primarily constrained through NH$_{3}$ observations \citep{rosolowsky:08}, with the exception of Per615, for which the dust temperature measured by \textit{Herschel} is adopted (see \citealt{scibelli:24}). For the Monte Carlo simulations, a relative uncertainty of 10\% is assumed for both $n_{\rm H_2}$ and $T_{\rm kin}$, meaning each value is randomly varied by 10\% around its nominal value. For molecules HCS$^{+}$, H$_{2}$CS and SO, a single line is fitted. In the case of OCS, when the two targeted transitions are detected, they are fitted simultaneously to compute the column density. In the case that only one transition is detected the column density is obtained just with that transition. For each molecule, the column density was determined by finding the model that minimizes the $\chi^{2}$ statistic. To quantify the associated uncertainties, a Monte Carlo method is employed: the observed line parameters, along with $n_{\rm H_2}$ and $T_{\rm kin}$, are randomly varied within their error ranges, and the RADEX fitting procedure is repeated for each trial.\

\begin{figure}[h]
\centering
\includegraphics[width=8cm]{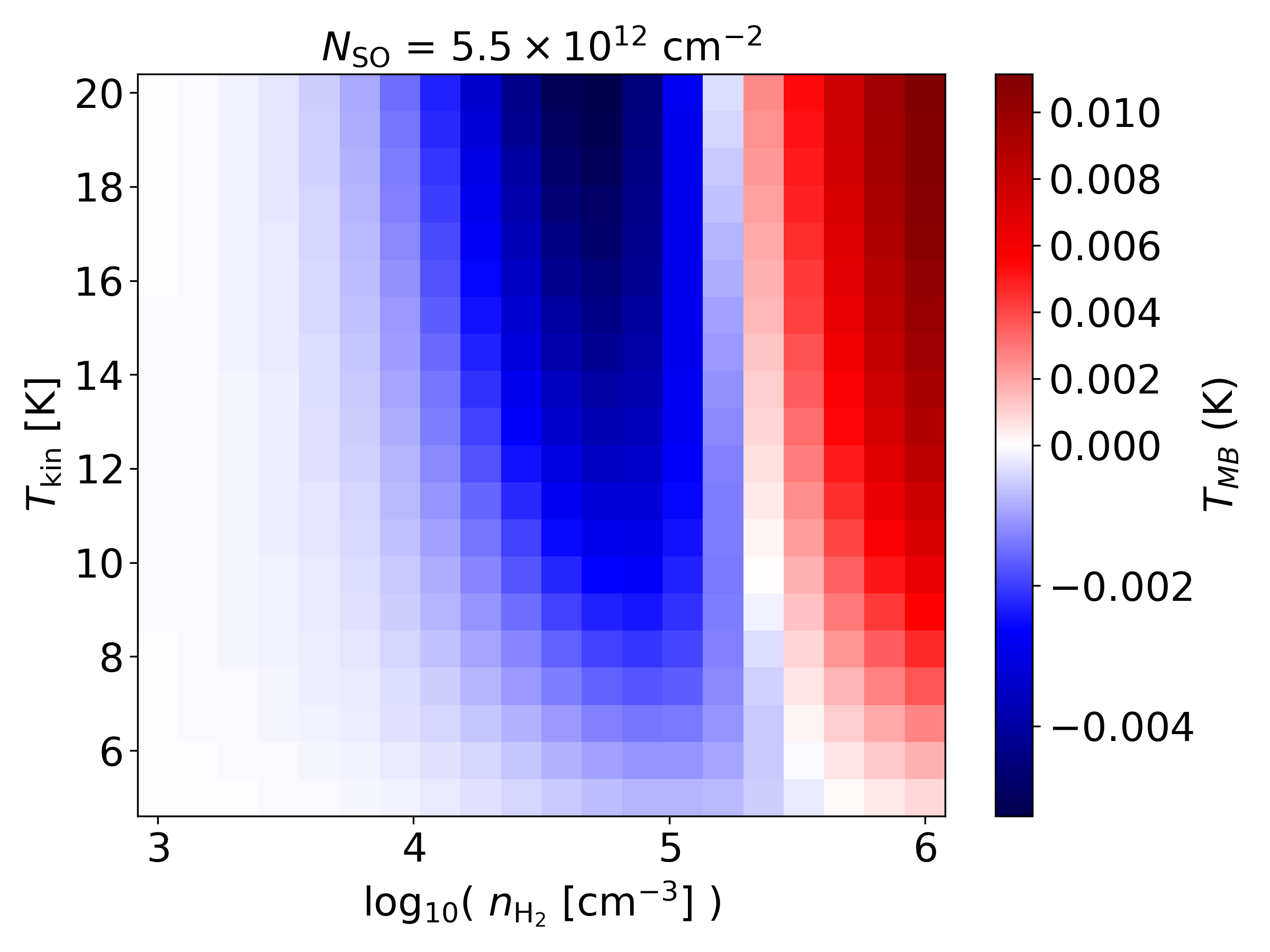}
\caption{Computed SO 3$_{2}$--2$_{2}$ $T_{\rm MB}$ with RADEX for an input $N_{\rm SO}$ = 5.5$\times$10$^{12}$ cm$^{-2}$ for a grid of $n_{\rm H_2}$ and $T_{\rm kin}$ input values. Red in the colour scale indicates a positive $T_{\rm MB}$, emission, and blue indicates negative $T_{\rm MB}$, absorption.}
\label{sogrid}
\end{figure}

We report the column densities of HCS$^{+}$, OCS and H$_{2}$CS computed with RADEX in Table \ref{coldenstrad}. The software was not able to find solutions for the SO 3$_{2}$--2$_{2}$ observed transition for any of the cores as the modelled transition by RADEX resulted in all cases in a negative $T_{\rm MB}$. To test whether the assumed $n_{\rm H_2}$ and $T_{\rm kin}$ input values are the issue, we ran RADEX for a grid of $n_{\rm H_2}$, $T_{\rm kin}$ and $N_{\rm SO}$ with 1$\times$10$^{3}$--1$\times$10$^{6}$ cm$^{-3}$, 5--20 K and 1$\times$10$^{11}$--1$\times$10$^{14}$ cm$^{-2}$, respectively. The only combinations that yield a $T_{\rm MB}$ of the order of the weakest SO detection (Per413, $T_{\rm MB}$= 9 mK) have a column density of $\geq$ 5.5$\times$10$^{12}$ cm$^{-2}$. Nevertheless, the modelled line brightness is always negative for $n_{\rm H_2}$ $<$ 3.2$\times$10$^{5}$ cm$^{-3}$. As an example we show the $T_{\rm MB}$ outputs for the $T_{\rm kin}$ vs $n_{\rm H_2}$ grid for a $N_{\rm SO}$ = 5.5$\times$10$^{12}$ cm$^{-2}$ in Figure \ref{sogrid}. This points to either the SO 3$_{2}$--2$_{2}$ transition tracing particularly dense gas, or an issue with the treatment of this transition by RADEX. If a test $n_{\rm H_2}$ of 5$\times$10$^{5}$ cm$^{-3}$ is adopted then the resulting SO column densities found for the core sample are higher by factors of 6 to 19 compared to the column densities derived assuming LTE in Section \ref{so}. If indeed SO is tracing higher-density regions within the cores, the strong direct correlation found with $n_{\rm H_2}$ is reinforced. Moreover, in this scenario, the abundance comparison with other molecules would not be completely fair as they would not be tracing the same material, but some of the more general conclusions may still hold. Nevertheless, drawing robust conclusions on whether the SO is indeed tracing denser material would require the observation and analysis of additional SO transitions.\

For some of the cores the $T_{\rm {ex}}$ of HCS$^{+}$ and H$_{2}$CS results in unphysical values, such as anomalous high values that go over a factor of a few the assumed T$_{\rm {kin}}$ or negative values. \cite{vandertak:07b} discusses the possibility of artifact population inversions for certain molecules and certain physical properties which lead to negative $T_{\rm {ex}}$ and $\tau$. Moreover, it is not uncommon that $T_{\rm {ex}}$ and $\tau$ are not constrained when only one transition is fitted. The results for each molecule as well as the presumed artifacts are discussed in the following sections.\ 

\begin{table}[H]
\begin{center}
\caption{Column densities calculated with the non-LTE RADEX software. }

\resizebox{8.5cm}{!}{%
\begin{tabular}{ lccc } 
\hline\hline
Core & HCS$^{+}$ & OCS & H$_{2}$CS   \\
$\#$ & & & \\
\hline
264 & (2.2 $\pm$ 0.3)$\times10^{11}$ & (3.2 $\pm$ 0.5)$\times10^{12}$ & (4.2 $\pm$ 0.3)$\times10^{12}$ \\
317 & (3.6 $\pm$ 0.5)$\times10^{11}$ & (3.5 $\pm$ 0.6)$\times10^{12}$ & (3.8 $\pm$ 0.3)$\times10^{12}$  \\
321 & (4.2 $\pm$ 0.4)$\times10^{11}$ & (5.2 $\pm$ 0.4)$\times10^{12}$ & (5.5 $\pm$ 0.3)$\times10^{12}$ \\
326 & (1.6 $\pm$ 0.1)$\times10^{12}$ & (2.4 $\pm$ 0.1)$\times10^{13}$ & (2.2 $\pm$ 0.1)$\times10^{13}$  \\
413 & (5.6 $\pm$ 0.3)$\times10^{11}$ & (2.8 $\pm$ 0.4)$\times10^{12}$ & (6.9 $\pm$ 0.4)$\times10^{12}$   \\
504 & (4.7 $\pm$ 0.3)$\times10^{11}$ & (5.3 $\pm$ 0.3)$\times10^{12}$ & (6.7 $\pm$ 0.3)$\times10^{12}$   \\
615 & (3.2 $\pm$ 0.3)$\times10^{11}$ & (3.0 $\pm$ 0.4)$\times10^{12}$ & (5.0 $\pm$ 0.3)$\times10^{12}$   \\
627 & (2.9 $\pm$ 0.3)$\times10^{11}$ & (1.7 $\pm$ 0.2)$\times10^{12}$ & (2.8 $\pm$ 0.2)$\times10^{12}$  \\
709 & (6.4 $\pm$ 0.7)$\times10^{11}$ & (1.9 $\pm$ 0.2)$\times10^{12}$ & (2.1 $\pm$ 0.2)$\times10^{12}$  \\
715 & (4.3 $\pm$ 0.5)$\times10^{11}$ & $<$3.0$\times10^{12}$ & (3.1 $\pm$ 0.2)$\times10^{12}$  \\
746 & (4.8 $\pm$ 0.5)$\times10^{11}$ & (2.6 $\pm$ 0.6)$\times10^{12}$ & (4.0 $\pm$ 0.3)$\times10^{12}$   \\
752 & (1.2 $\pm$ 0.2)$\times10^{11}$ & (8.0 $\pm$ 0.6)$\times10^{12}$ & (8.3 $\pm$ 1.3)$\times10^{11}$   \\
768 & (3.4 $\pm$ 0.3)$\times10^{11}$ & (1.7 $\pm$ 0.4)$\times10^{12}$ & (1.9 $\pm$ 0.1)$\times10^{12}$ \\
799 & (3.7 $\pm$ 0.5)$\times10^{11}$ & (3.5 $\pm$ 0.2)$\times10^{12}$ & (3.7 $\pm$ 0.1)$\times10^{12}$   \\
800 & (3.8 $\pm$ 0.3)$\times10^{11}$ & (7.1 $\pm$ 0.5)$\times10^{12}$ & (6.7 $\pm$ 0.3)$\times10^{12}$ \\
\hline
\textbf{Median} & (3.9 $\pm$ 0.5)$\times10^{11}$ & (3.4 $\pm$ 0.8)$\times10^{12}$ & (3.9 $\pm$ 1.0)$\times10^{12}$ \\
\textbf{Deviation from LTE} & 1 - 1.8 & 1 & 1 - 1.5 \\
\hline
\end{tabular}%
}
\tablefoot{The upper limits correspond to the 3$\sigma$ level. The median molecular column densities are reported at the bottom, including all of the cores with detections. The median uncertainty corresponds to the 16th and 84th percentiles, equivalent to a 1-sigma confidence interval, derived via Monte Carlo simulations. Additionally, the deviation factor range from the LTE column densities shown in Table \ref{coldenst} are shown. A factor of 1 means that the non-LTE and LTE column densities agree.  }
\label{coldenstrad}
\end{center}
\end{table}

\subsection{HCS$^{+}$}

The HCS$^{+}$ column density computed with RADEX ranges in the sample from (1.2 $\pm$ 0.2)$\times10^{11}$ cm$^{-2}$ (Per752) to (1.6 $\pm$ 0.1)$\times10^{12}$ cm$^{-2}$ (Per326) with a median of (3.9 $\pm$ 0.5)$\times10^{11}$ cm$^{-2}$ (Table \ref{coldenstrad}). The column densities calculated with the LTE and non-LTE (RADEX) methods agree within uncertainties for cores Per264, Per317, Per321, Per326, Per709 and Per715. The values found with the two methods differ for cores Per413, Per504, Per615, Per627, Per746, Per752, Per768, Per799 and Per800 by factors between 1.4 and 1.8.\

The $T_{\rm {ex}}$ for one core, Per752, has a value of -560 K. The $T_{\rm {ex}}$ for the rest of the cores ranges from 13 to 96 K with a median of 19 K and an average of 30 K. Three of the cores with remarkably high $T_{\rm {ex}}$ are Per709, Per627 and Per715 with values 35, 91 and 96 K. The median $\tau$ is of 0.002, excluding the one for Per752, which also results in a negative value. As only one HCS$^{+}$ transition is fitted, the obtained $T_{\rm {ex}}$ and $\tau$ values are not constrained, which can explain the odd values obtained.\

\subsection{OCS}

For all the cores the OCS column density computed with RADEX was calculated by fitting the two transitions simultaneously. For cores Per746 and Per768 the column density was only calculated with one transition, due to only one transition being detected towards these cores. Lastly, for core Per715, we give an upper limit (Table \ref{coldenstrad}). The column density of OCS ranges from (1.7 $\pm$ 0.2)$\times10^{12}$ cm$^{-2}$ (Per627) and (2.4 $\pm$ 0.1)$\times10^{13}$ cm$^{-2}$ (Per326) with a median of (3.4 $\pm$ 0.8)$\times10^{12}$ cm$^{-2}$. The column densities computed with the LTE and non-LTE (RADEX) methods agree within errors for all cores.\

The $T_{\rm {ex}}$ for the 3--2 transition ranges from 9 to 15 K with a median of 11 K, and for the 4--3 transition it ranges from 10 to 15 K with a median of 12 K. The median $\tau$ for the 3--2 transition is 0.003 and for the 4--3 transition is 0.005. In the case of OCS no anomalous results are seen. This can be due to the simultaneous fit of two transitions for most of the cores constraining $T_{\rm {ex}}$ and $\tau$ better than for the other molecules.\

\begin{figure}[h]
\centering
\includegraphics[width=8cm]{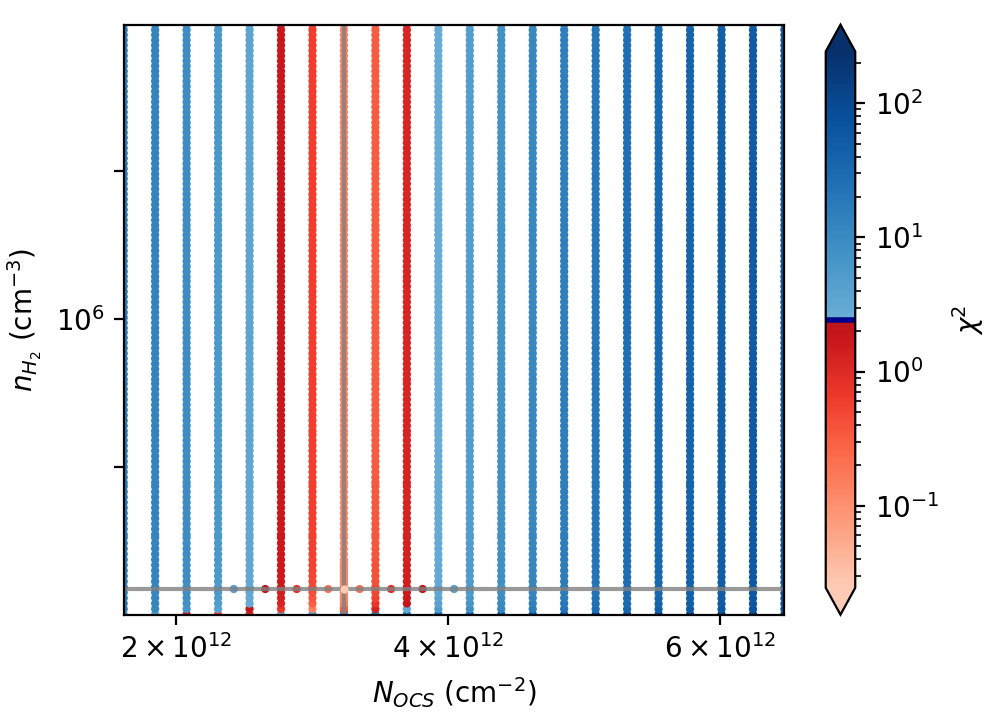}
\caption{The agreement between the observations and RADEX $\chi$$^{2}$ values, is plotted in colour on a grid of OCS column density and H$_{2}$ volume density. The change between red and blue is set by the 1$\sigma$ level due to the observed rms level. The gray vertical and horizontal lines correspond to the ``best model''. }
\label{ocsgrid}
\end{figure}

For the cores with detections of both OCS transitions, we additionally try to constrain the $n_{\rm H_2}$ and $N_{\rm H_2}$ simultaneously by fixing the $T_{\rm {ex}}$ using the values in Table \ref{source}. As only two transitions are available, constraining also the $T_{\rm {ex}}$ is unfeasible. The $n_{\rm H_2}$ is degenerate and not constrained for any of the sources. As an example, the exploration of the $n_{\rm H_2}$ and $N_{\rm H_2}$ grid for core Per264, done with the \textit{RADEX Line Fitter} \footnote{\url{https://github.com/andresmegias/radex-line-fitter}} software (Figure \ref{ocsgrid}). The reason why the volume density may not be constrained is that the OCS transitions are in LTE conditions. The critical densities ($n_{\rm crit}$ = $A_{ij}$ / $k_{ij}$($T$), where $k_{ij}$($T$) is the collisional coefficient rate at a temperature $T$) of the 3--2 and 4--3 transitions between 10 and 15 K are $\sim$1.8$\times$10$^{3}$ cm$^{-3}$ and $\sim$1.7$\times$10$^{3}$ cm$^{-3}$. This critical densities are lower than the volume densities presented in Table \ref{source}, which points at OCS being at LTE. That would also explain the agreement between the LTE calculated column densities and the non-LTE column densities derived with RADEX. Another factor that could be contributing to the inability to constrain $n_{\rm H_2}$ is that the OCS transitions are close in $E_{up}$.\

\subsection{H$_{2}$CS}

The total H$_{2}$CS column densities are computed from fitting the para-H$_{2}$CS 1$_{0,1}$--0$_{0,0}$, and by assuming the statistical ortho-to-para ratio of 3. In the sample of cores targeted in this study the column density ranges from (8.3 $\pm$ 1.3)$\times10^{11}$ cm$^{-2}$ (Per752) and (2.2 $\pm$ 0.1)$\times10^{13}$ cm$^{-2}$ (Per326) with a median of (3.9 $\pm$ 1.0)$\times10^{12}$ cm$^{-2}$. The column densities computed with the LTE and non-LTE methods agree for cores Per264, Per317, Per413, Per504, Per615, Per627, Per709, Per715, Per746, Per752, Per768 and Per800. For cores Per321, Per326 and Per799 the column densities disagree by factors between 1.3 and 1.5, slightly less than the disagreement of the HCS$^{+}$ column densities.\

As for HCS$^{+}$, Per752 has a negative $T_{\rm {ex}}$ value of -66 K. The cores Per627, Per709, Per746, Per799 and Per800 show abnormally large $T_{\rm {ex}}$: 3941, 202, 49, 121 and 221 K, respectively. For the rest of the sources the $T_{\rm {ex}}$ median is 19 K and the average is 22 K, which are still slightly larger than the assumed T$_{\rm {kin}}$ of the cores. The $\tau$ for cores Per627, Per715 and Per752 are negative. For the rest of the source sample the $\tau$ has a median of 0.003. As for HCS$^{+}$, only one H$_{2}$CS transition is fitted with RADEX, which can be the reason for the unphysical $T_{\rm {ex}}$ and $\tau$ values derived. \

\section{Outliers in $n_{\rm H_2}$ and T$_{\rm {kin}}$ correlations}\label{physa}

\begin{figure}[h]
\centering
\includegraphics[width=8cm]{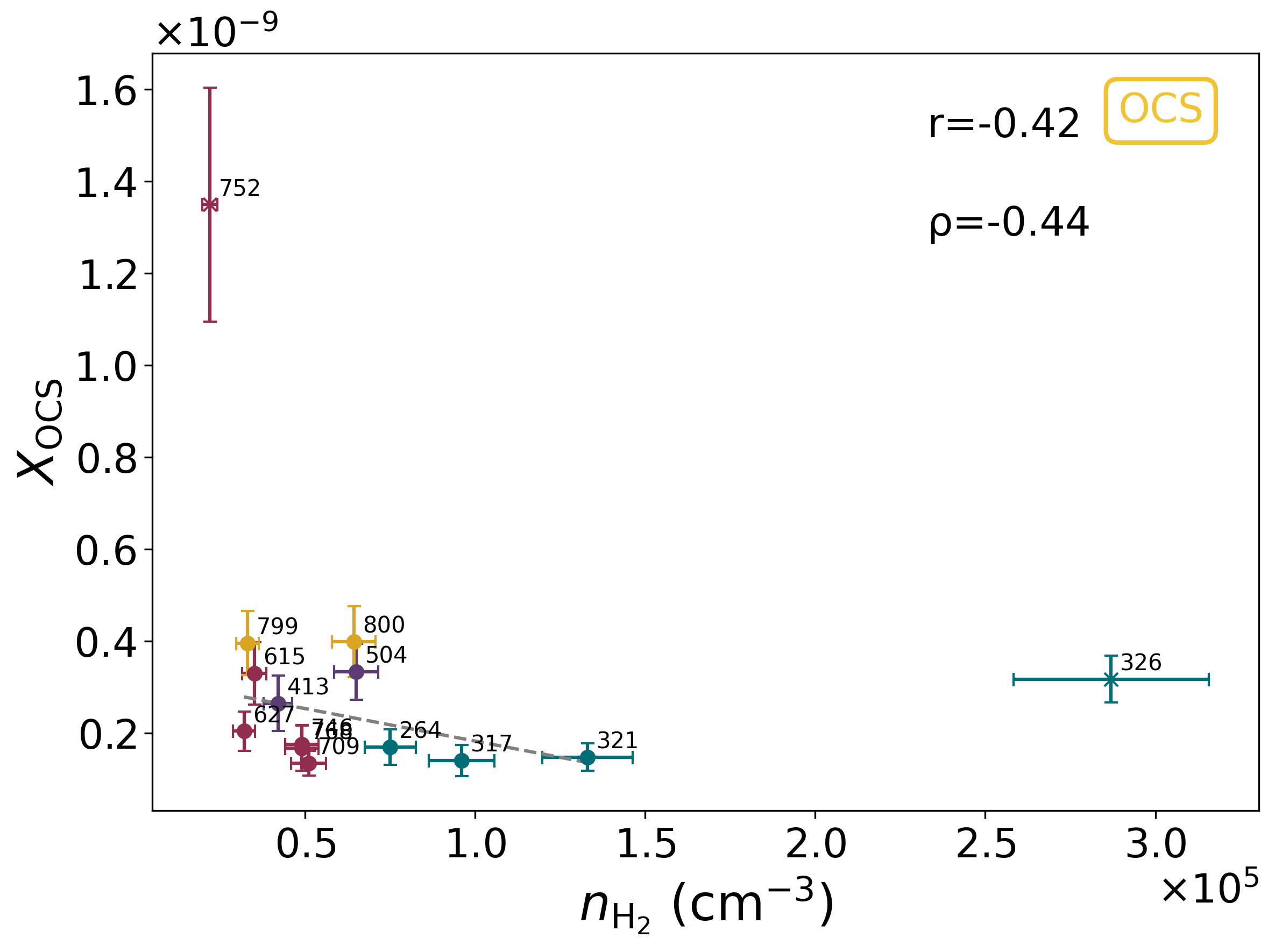}
\caption{Abundance correlation of OCS with $n_{\rm H_2}$ with the Pearson and the Spearman correlation coefficients, $r$ and $\rho$, plotted on the top right corner. The dashed gray line represents the Pearson correlation. Per326 and Per752 are outliers, and thus excluded from the correlation and plotted with a cross symbol.  }
\label{ocsphys}
\end{figure}

\begin{figure*}[h]
\centering
\includegraphics[width=\textwidth]{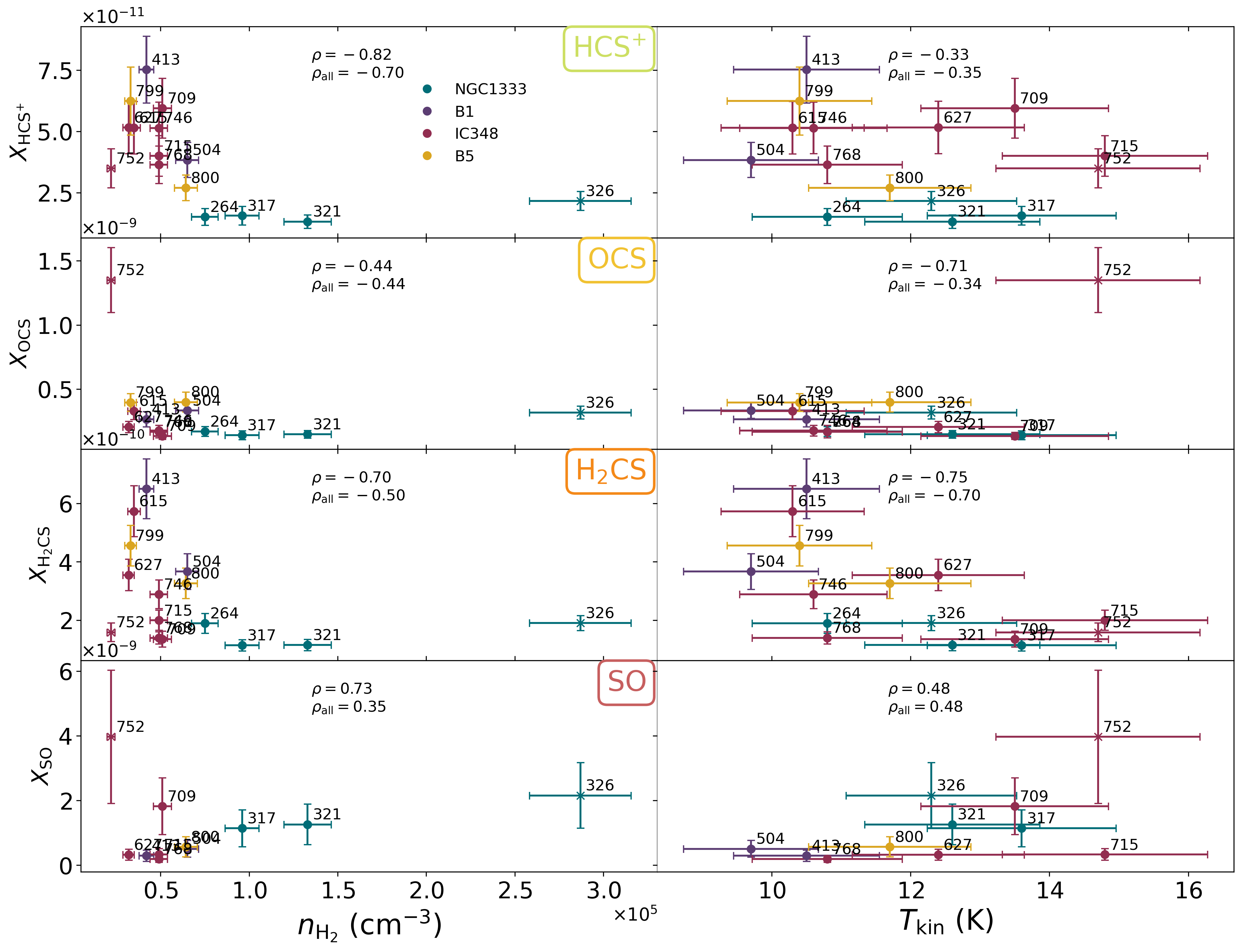}
\caption{Abundances of HCS$^{+}$, OCS, H$_{2}$CS and SO plotted against $n_{\rm H_2}$ and $T_{\rm kin}$ in the first and second columns, respectively. Spearman correlation coefficients for the sample excluding core outliers ($\rho$), as well as for the full core sample ($\rho_{all}$) are printed on the top centre. The cores belonging to different regions within the Perseus molecular cloud are plotted with different colours (NG31333 in teal, B1 in purple, IC348 in red, and B5 in yellow). Per326 and Per752 are outliers and therefore excluded and plotted with a cross symbol.}
\label{phys}
\end{figure*}

To show the identification and effect of the outlier abundances of Per326 and Per752 in the correlations, the Pearson and Spearman correlations for OCS vs $n_{\rm H_2}$ are presented in Figure \ref{ocsphys}. When including both Per326 and Per752 the $r$ and $\rho$ values are -0.20 and -0.44, respectively. The factor two difference between the Pearson and Spearman coefficients, hint at the abundance of OCS at Per326 and Per752 being outliers. The Per326 appears as an outlier in the correlations with $n_{\rm H_2}$ due to its significantly higher column density value compared to the rest of the sample (see Section \ref{discussion} for more details). In the case of Per752 apart from OCS, the other abundance that appears to be an outlier is the SO one. This behaviour seems to be directly related to the OCS and SO abundances towards Per752 (see Section \ref{discussion} for more details). Thus, the abundances for all molecules derived towards Per326, as well as the abundances of OCS and SO towards Per752 are excluded from the analysis. These exclusions result in closer $r$ and $\rho$ coefficients, -0.42 and -0.44, respectively. Due to the nature of the data the Spearman rank coefficient is chosen to present correlations in this study (Figure \ref{phys}). \

\section{Additional material for the molecular abundance comparison}\label{molcompa}

\begin{figure*}[h]
\centering
\includegraphics[width=22cm]{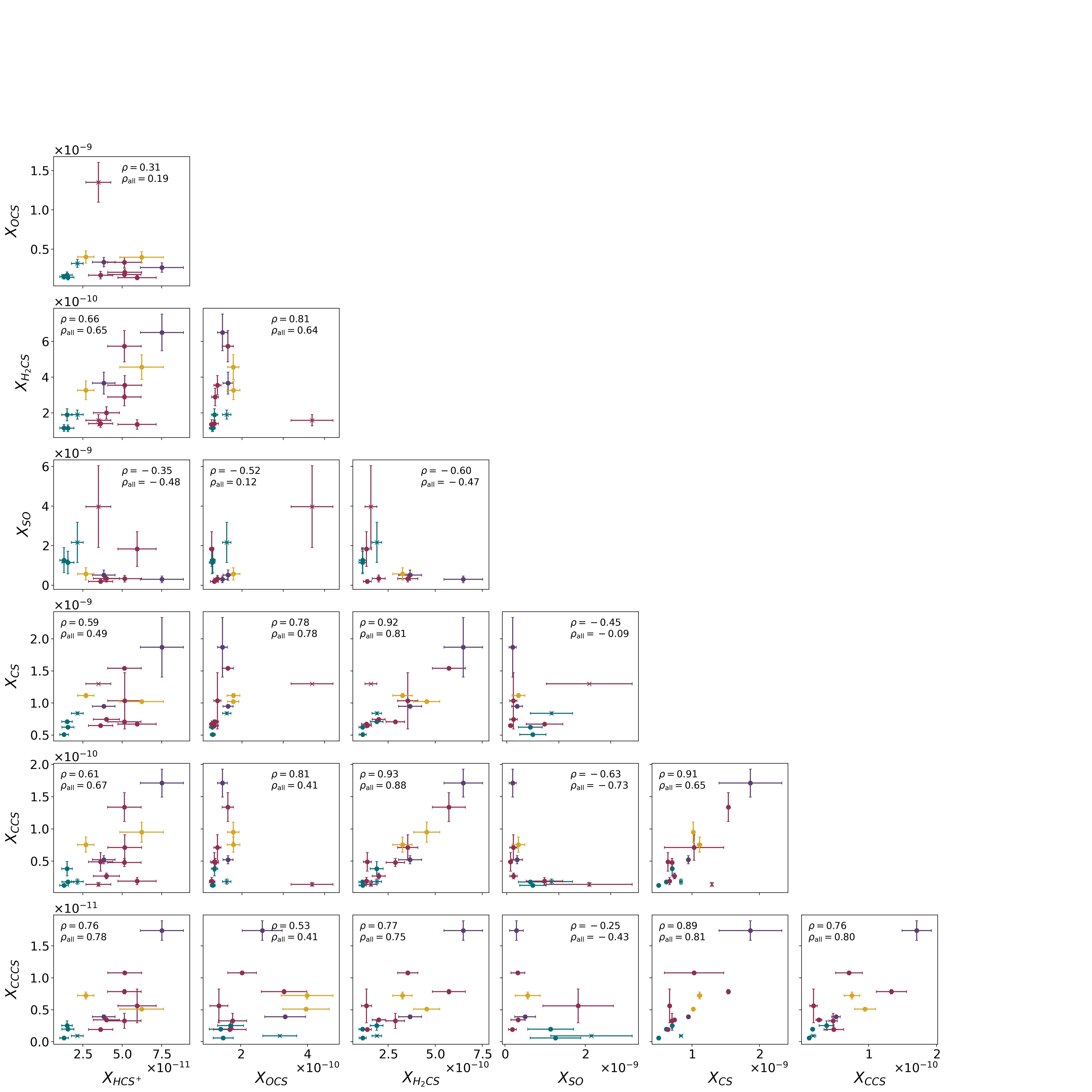}
\caption{Pair molecular abundance comparisons. Spearman correlation coefficients are printed. Additionally, for pairs with outliers, the spearman correlation correlation coefficient including all of the cores ($\rho_{all}$), is plotted. The cores belonging to different regions within the Perseus molecular cloud are plotted with different colours (NG31333 in teal, B1 in purple, IC348 in red, and B5 in yellow). Per326 and Per752 are outliers and are plotted with a cross symbol instead of a circle.}
\label{triangle}
\end{figure*}

Similarly to the molecular abundance vs $n_{\rm H_2}$ and $T_{\rm kin}$ tests, both the Pearson and the Spearman rank methods are used in the analysis of the correlations between molecular abundances. The molecular abundances for Per326 do not show outlier behaviour in these comparisons. The reason why these points dominate the correlations with $n_{\rm H_2}$ is due to the large Per326 $n_{\rm H_2}$ value compared to the rest of the cores. Nevertheless, for consistency these are excluded also from the molecular abundance correlations. In the molecular abundance comparison pairs the OCS and SO Per752 appear as clear outliers, and as such, they are excluded from all the analyses. The Pearson and Spearman correlation coefficients after the exclusions agree within 0.05. Other points could be also classified as outliers, but as these cases are not as clear as for the one of Per752 we include them and provide the correlation degrees as measured by the Spearman rank coefficient, as this has a better treatment of outliers compared to the Pearson correlation coefficient. The uncertainties of some molecular abundances towards particular cores may have an influence on the derived correlations. All of the molecular abundance comparisons are plotted in Figure \ref{triangle}.   \

\end{document}